\documentclass[12pt,preprint]{aastex}

\usepackage[parfill]{parskip}    % Activate to begin paragraphs with an empty line rather than an indent
\usepackage{graphicx}
\usepackage{float}
\usepackage{color}
\usepackage{url}
\usepackage{threeparttable}
\def\arcsecpoint{$''\!.$}
\def\deg{$^{\circ}$}
\shorttitle{X-ray Emission-lines in NGC\,1068}
\shortauthors{Kraemer et al.}

\begin{document}

\title{Physical Conditions in the X-ray Emission-line Gas in NGC\,1068}

\author{S. B. Kraemer \altaffilmark{1}, N. Sharma \altaffilmark{2}, T. J. Turner \altaffilmark{2}, Ian M. George \altaffilmark{2}}

\and 

\author{ D. Michael Crenshaw \altaffilmark{3} }

\altaffiltext{1}{Institute for Astrophysics and Computational Sciences, Department of Physics, 
The Catholic University of America, 
Washington, DC 20064}

\altaffiltext{2}{Department of Physics, University of Maryland Baltimore County, 
   Baltimore, MD 21250, U.S.A}

\altaffiltext{3}{Department of Physics and Astronomy, Georgia State University, Astronomy Offices, One Park Place South SE, Suite 700, Atlanta, GA 30303}

\begin{abstract}
We present a detailed, photoionization modeling analysis of {\it XMM-Newton}/Reflection Grating Spectrometer observations
of the Seyfert 2 galaxy NGC~1068. The spectrum, previously analyzed by \citet{kinkhabwala2002},
reveals a myriad of soft-Xray emission lines, including those from H- and He-like carbon, nitrogen,
oxygen, and neon, and M- and L-shell iron. As noted in the earlier analysis, based on the narrowness
of the radiative recombination continua,  the electron temperatures in the emission-line gas
are consistent with photoionization, rather than collisional ionization. 
The strengths of the carbon and nitrogen emission lines, relative to those of oxygen,
suggest unusual elemental abundances, which we attribute to star-formation history of the
host galaxy. Overall, the emission-lines are blue-shifted with respect to systemic, with radial velocities
$\sim$ 160 km s$^{-1}$, similar to that of [O~{\sc iii}] $\lambda$5007, and thus consistent with
the kinematics and orientation of the optical emission-line gas and, hence, likely part of an  AGN-driven
outflow. We were able to achieve an acceptable fit
to most of the strong emission-lines with a two-component photoionization model, generated with Cloudy. The two
components have ionization parameters and column densities of log$U = -0.05$ and 1.23, and log$N_{\rm H} =
20.85$ and 21.2, and covering factors of 0.35 and 0.84, respectively. The total mass of the X-ray
gas is roughly of an order of magnitude greater than the mass of ionized gas determined from optical
and near-IR spectroscopy, which indicates that it may be the dominant component
of the narrow line region.
Furthermore, we suggest that the medium which produces the scattered/polarized optical emission in
NGC~1068 possesses similar physical characteristics to those of the more highly-ionized of the X-ray model components.

\end{abstract}

\keywords{galaxies: active - galaxies: individual: NGC~1068 - galaxies: Seyfert - X-rays: galaxies}

\section {Introduction}
\label{Introduction}

Active Galactic Nuclei (AGN) of which Seyfert galaxies are relatively low luminosity $L_{bol}$ $\lesssim$ several $\times$ 10$^{45}$ ergs s$^{-1}$)
examples in the local Universe($z \lesssim 0.1$), are thought to be powered by accretion of matter onto
super-massive black holes, which reside at the gravitational centers of the host galaxies.
The properties of Seyfert galaxies and how these are understood within the context of the unified model
\citep{antonucci1993} have been discussed in several of our previous papers \citep[e.g.,][]{kraemer2009}.
However, it is useful in the context of this paper to mention the scales of the various regions that contribute to the 
spectra of Seyferts. The continuum source, which is $\sim$ light hours in extent \citep[e.g.,][]{edelson1996} and surrounding broad emission-line region, which
ranges in size from several to tens of light days \citep[e.g.,][]{peterson2004}, are only directly 
observable in Seyfert 1 galaxies. In contrast, the narrow-line region (NLR), which is comprised of lower-density gas in which 
the forbidden lines and narrower components
of the permitted lines form, can extend out to $\sim$ 1 kpc \citep[e.g.,][]{pogge1988}.

Due to its proximity,  the Seyfert 2 nucleus of the 
barred-spiral galaxy NGC 1068 \citep{bland1997} 
is the most extensively studied such object to date. 
Redshift-independent distance estimates place the
host galaxy between ~10 and ~16 
Mpc away \citep[e.g.][]{tully2009, sofue1991}. To be consistent 
with the bulk of the current literature here we adopt the mean value
of
12.65 Mpc from NED\footnote{The NASA/IPAC Extragalactic Database 
(NED) is operated by the Jet Propulsion Laboratory, California 
Institute of Technology, under contract with the National Aeronautics 
and Space Administration}, such that 1\arcsec~is roughly equivalent to 60 pc.

Ground-based narrowband images of the NLR in NGC~1068 show that the optical
emission-line gas possesses a biconical morphology, roughly parallel to the
major axis of the radio emission, extended northeast and southwest of the
nucleus \citep{pogge1988}. Narrow-band optical images obtained with the {\it Hubble
Space Telscope (HST)} revealed that the inner NLR also possesses a biconical
morphology \citep[e.g.,][]{evans1991} and is comprised of numerous 
knots and filaments. The hidden AGN is thought to be 0\arcsecpoint3 south 
of the optical continuum peak \citep{capetti1997}, AKA the ``Hot Spot''
\citep{kraemer&crenshaw2000}.

Optical and UV spectra of NGC~1068, in particular those obtained
with the Space Telescope Imaging Spectrograph abroad {\it HST},
 have been studied in detail \citep[e.g.,][]{crenshaw&kraemer2000A,
 crenshaw&kraemer2000, kraemer&crenshaw2000C, kraemer&crenshaw2000}.
 The optical emission is dominated by the Hot Spot, spectra of which
 show emission lines from an extreme range of ionization, e.g., 
 [O~{\sc i}]~$\lambda$6300 to~[Fe~{\sc xiv}] $\lambda$5303 and [S~{\sc xii}]~$\lambda$7611; the latter
 two are likely the footprint of the X-ray emission line gas.
 The Hot Spot is also a source of strong scattered continuum radiation. Within
 $\sim$200 pc, on either side of the nucleus, the gas is somewhat less highly
 ionized, however strong N~{\sc v}~$\lambda$1240, C~{\sc iv} $\lambda$1550, and 
 [Ne~{\sc v}]~$\lambda$3426 are still present. Photoionization modeling has shown that the
 gas is ionized by the continuum radiation from the central source, although
 there are localized regions which show evidence for enhanced ionization
 due to shocks.
  
 Both red- and blue-shifted emission lines are present on either side
of the nucleus \citep{crenshaw&kraemer2000}. Based on kinematic modeling
\citep{vdas2006}, the observed velocities result from 
outflow along a hollow bicone, with a half opening angle
of 40\deg. The axis of the bicone is inclined
5\deg~ with respect to the plane of the sky, with the NE side pointed
towards the observer, which is consistent with the morphology
of the H~{\sc i} 21 cm emission \citep{gallimore1994}. The maximum de-projected outflow
velocity is $\sim$ 2000 km  s$^{-1}$ with a peak $\sim$ 140 pc from the
central nucleus, after which there is rapid deceleration to systemic.
Although outflowing, photoionized gas is
suggestive of radiative acceleration, the velocity profiles are not
fully consistent with any simple form of AGN-driven flow \citep[]{vdas2007, everett2007}. The main issue is that
the flow appears to accelerate much more gradually than predicted.
One possibility is mass loading, perhaps in the form of interaction with
more tenuous ambient gas.     

Absorption of the 2--10 keV X-ray continuum in Type 1 AGN by photoionized
gas (now generally referred to as ``warm absorbers'') was first suggested 
by \citet{halpern1984}. Although modification of the X-ray continuum
by the absorber could account for the overall spectral properties of such
sources, {\it ASCA} spectra of Seyfert 1 galaxies revealed evidence for an 
unabsorbed component 
of emission \citep{george1998}, particularly in the soft X-ray band
(energies $<$ 1 keV).  In their analysis of
these, \citet{george1998} found that, in many cases, the fit statistics improved
when a component of unabsorbed emission-line gas was included. In fact,
\citet[]{netzer1993, netzer1996} had predicted that there would be strong 
soft-X-ray emission lines associated with warm absorbers.

The unprecedented X-ray spectral  resolution afforded by the 
 {\it XMM-Newton}  (hereafter {\it XMM}) Reflection Grating Spectrometer (RGS) and the {\it Chandra}/ High Energy Transmission Grating (HETG) and Low Energy Transmission Grating (LETG)
  have revealed the myriad of emission lines predicted by \citet{netzer1993}. Furthermore, based on {\it Chandra}/HETG observations,
  \citet{ogle2000} found extended soft X-ray emission in the Seyfert 1 galaxy, NGC~4151. As it is roughly spatially
  coincident with the optical [O~III] emission and radio continuum, it likely arises in the NLR. Similarly, the NLR
is the source of the soft X-ray emission in Seyfert 2 galaxies. For example,  
X-ray observations of NGC~1068 show an extended region of soft X-ray emitting gas, broadly coincident with the [O {\sc iii}] bicone \citep{young2001,ogle2003}. 
The extended emission component likely has contributions from both line emission and a small electron-scattered component of the X-ray continuum.   

 Previous analyses, based upon RGS \citep{kinkhabwala2002} and {\it Chandra}/LETG \citep{brinkman2002} and HETG \citep{ogle2003} observations  have revealed 
a multitude of emission lines in NGC~1068. 
Particularly prominent are  strong H- and He-like lines of C, N and O. These are accompanied by weaker highly-ionized lines from Ne, Mg, Si, S and Fe. 
Via its superior spatial resolution, the {\it Chandra} observations revealed two peaks of X-ray emission, one centered on the nucleus and the
other 3$^{''}$ - 4$^{''}$ to the NE. The overall emission-line spectra was found to be similar in both regions
\citep{brinkman2002, ogle2003}.  \citet{ogle2003} suggested that the
X-ray emission-line was the source of the scattered optical continuum, detected
via spectro-polarimetry \citep{miller1991}. 
Consideration of the ensemble of line measurements has suggested that both 
photoionization and photoexcitation are important ionization processes for the X-ray emitter \citep{kinkhabwala2002, ogle2003}, with no
clear evidence of a collisionally ionized component. However, none of these authors generated detailed 
photoionization models to analyze the physical properties of the emission-line gas.
 
More recently, \citet{kallman2014} analyzed a 450 ksec {\it Chandra}/HETG spectrum of NGC~1068,
using photoionization models generated with the code {\sc xstar} \citep{kallman2004}. Overall, their
results support the claims in \citet{kinkhabwala2002} and \citet{ogle2003}, specifically that the emission-line 
gas is photoionized, with some evidence for photoexcitation, and the emission-lines are blue-shifted, indicative of
outflows. They found that multiple zones, characterized by a range of 
ionization parameters and column densities were required to match the measured
emission line fluxes. Furthermore, in order to fit the spectra, they
allowed the oxygen and iron abundances to vary among the components. However,
{\it Chandra}/Medium Energy Grating has a low effective area at wavelengths $>$ 25 \AA~. As
a result, \citet{kallman2014} were unable to compare the oxygen lines to many of the strong lines of nitrogen
and none of carbon, and therefore they were unable to determine how their assumed 
abundances might result within the constraints of nuceleosynthesis models.
We will revist these points in Section~\ref{Photoionization Modeling}.

In this paper we re-investigate the soft X-ray spectra obtained using the RGS in the 
context of photoionization models generated with {\sc cloudy} \citep{ferland1998}. The 
goals of our study include a reassessment of the physical processes responsible for ionizing 
the X-ray gas and a better understanding of the relationship between the X-ray emitter and other 
key components of the nuclear outflow.  Progress on these questions will, in turn, lead to a
 better understanding of the energy and material transport between the active nucleus and the host galaxy.

\section {Observations and Data Analysis}
\label{Observations and Data Analysis}

The observation of NGC 1068 reported here was made by {\it XMM} on 2000 July 29 - 30 (OBSIDs 011100101, 0111200201. 
The standard RGS and PN data products were extracted from the archive, having been produced by the {\it XMM} Science Analysis Software (SAS) v6.6.0. The RGS 
offers a useful bandpass ${\sim}$ 0.4 - 2.0 keV (${\sim}$ 6 - 31{\AA}).  Due to the failure of RGS1 CCD7 and RGS2 CCD4 (soon after launch), there are no useful 
data over the wavelength range 11-14 {\AA} (0.9-1.2 keV) and 20-24 {\AA}  (0.51-0.62 keV) for RGS1 and RGS2, respectively. Subsequent reprocessing and analysis were 
performed using a more recent version of SAS\footnote{\rm http://xmm.esac.esa.int/sas/} (v10.0.0) and various tasks 
from  the {\sc heasoft}\footnote{http://heasarc.gsfc.nasa.gov/docs/software/lheasoft.html} (v6.9) software suite. %(XMM-SOC-CAL-TN-0030).  

In order to determine the RGS wavelength scale the position of the zeroth order must be determined. This is not possible using the RGS instrument alone.
 Thus, we used the {\it HEAsoft} "xrtcentroid" task (within {\sc ximage} v0.2.9) to determine the centroid of the image from the co-aligned PN instrument 
 in the RGS wavelength band. The centroid position was found to be at RA = 02$^{\rm h}$42$^{\rm m}$40.70$^{\rm s}$, DEC = -0$^{\rm o}$00${\arcmin}$46.24${\arcsec}$ (J2000). 
 We estimated the uncertainty of our centroid position to be 1.5$\arcsec$ based upon repeated trials of the task, i.e., less than one image pixel (1.6$\arcsec$ on a side). 
  Our X-ray centroid position is consistent with that derived from {\it Chandra} observations of the source \citep{young2001} but is ${\sim}$ 6.3 ${\arcsec}$ away 
  from the position assumed in the (SAS v6.6.0) processing of the RGS data that created the archived data products. (The processing software uses the position supplied 
  by the principal investigator). We also note that our centroid position is ${\sim}$ 5${\arcsec}$ away from the X-ray centroid position determined by 
 \citet{kinkhabwala2002} for this {\it XMM} observation, attributable to improvements in
 attitude determination software between the original processing of the data and the current archived version. As the RGS energy 
 scale is determined by the angle a photon has been 
dispersed relative to the zeroth order position, this difference in attitude solution 
does not affect the line energy determination relative to that performed by  \citet{kinkhabwala2002}. 
Further to the uncertainty on photon energy related to the statistical uncertainty in determination 
of the centroid and therefore the dispersion angle, there is an additional uncertainty in the absolute energy scale that corresponds to a 
$1 \sigma$ uncertainty in velocity of 105 km s${^{-1}}$ at 20{\AA}.  

Given the above, we reprocessed the RGS data from both OBSIDs using SAS v10.0.0 to produce co-added source and background spectra for each RGS, based on our X-ray centroid position. The total exposure times were 84.7 ks and 82.5 ks for RGS 1 and 2 respectively. The mean source count rates (in the 0.4-2 keV band) were 0.568 ${\pm}$ 0.003 cts/s and 0.516 ${\pm}$ 0.003 cts/s  for RGS 1 and RGS 2 respectively. The background comprised 
$\sim 20\%$ of the total count rate.

\subsection{Spectral Analysis}
\label{Spectral Analysis}

\subsubsection{Initial Line Fitting}

\begin{figure}[ht]

\includegraphics[scale=0.85,viewport=120 0 600 500]{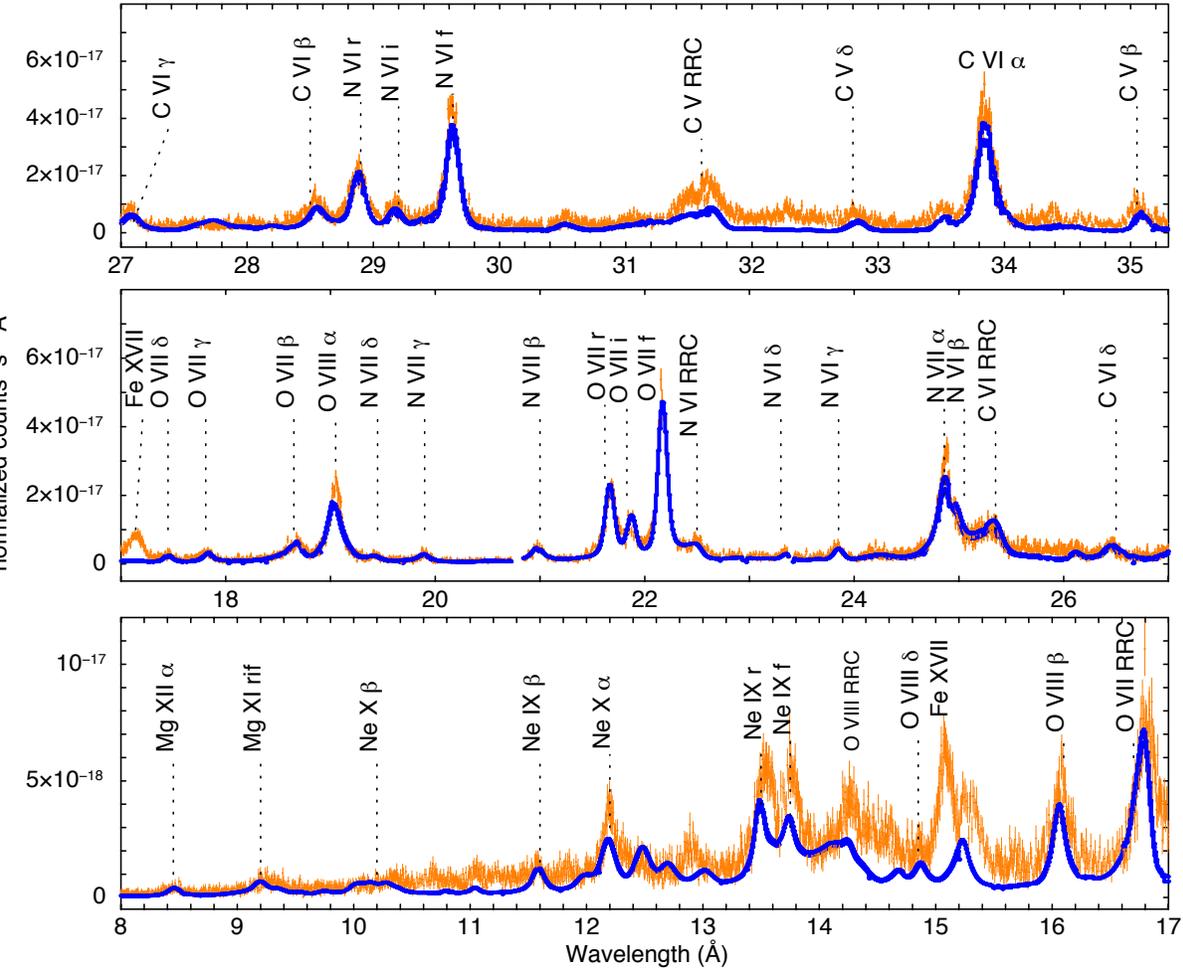}  
\vspace{-0.35in}
\caption{The total model comprising of both the zones (blue) compared to the
combined RGS1 and RGS2 spectrum (orange).
Note that the model produces a fairly good fit, except for $\lambda < 17 \AA$,
as discussed in Section~\ref{Final Fit}.}
\label{whole spectra}

\end{figure}

 The source spectral (without background subtraction), background and response files obtained after reprocessing the data (as discussed above), were loaded into {\sc xspec} \citep{arnaud2010} to create the background-subtracted source spectra from RGS 1 and 2,  as shown in Figure~\ref{whole spectra} ({\it orange} color). This methodology allows {\sc xspec} to perform the background subtraction and preserve the full statistical information from the data. 
The RGS spectra contained $>$ 10 counts in most of the channel range used allowing us to utilize the $\chi^{2}$ statistic for fitting.

Inspection of the spectra revealed several prominent lines (Figure~\ref{whole spectra}) including those from H-like and  He-like transitions of O, C, N, Ne, Mg and Si.  
Of particular interest are the very prominent triplets that arise from He-like species, such as N {\sc vi}, O {\sc vii} and Ne {\sc ix}, that can yield constraints on the gas density and excitation mechanism \citep{porquet&dubau2000}.

To extract the parameters for these lines, narrow sections of the data were fit using {\sc xspec}. As far as was possible, the lines were isolated (by ignoring data around the selected range) and fit individually such that each line could be fit accurately.  
Each line profile was modeled using a Gaussian component whose flux, width and energy were allowed to vary (Table~\ref{observations}).  The Galactic column density was accounted for in the spectral analysis by using the neutral absorption model {\sc tbabs}  \citep{wilms2000}. The column density for the {\sc tbabs} component was allowed to vary between 2.92 $\times 10{^{20}}$ cm${^{-2}}$ \citep{dickey1990} and  3.53 
$\times$ 10${^{20}}$ cm${^{-2}}$ \citep{kalberla2005}, to account for the uncertainty in this quantity.  The best-fit value of the Galactic column pegged at the high end of this 
range in the fit  and, therefore was initially fixed at  3.53 $\times$
10${^{20}}$ cm${^{-2}}$  (but see Section~\ref{Final Fit}). 
The continuum close to each line was fit with a power-law component
 that was allowed to vary.  The best fit line parameters are 
 detailed in Table~\ref{observations} for lines above the threshold of 
 observed flux $>$ 10${^{-6}}$ photons cm${^{-2}}$ s${^{-1}}$.  We 
 tabulated  1${\sigma}$ errors on each parameter 
 (i.e., calculated at 68\% confidence). 

\begin{figure}[ht]

\includegraphics[scale=0.66,viewport=5 0 500 540]{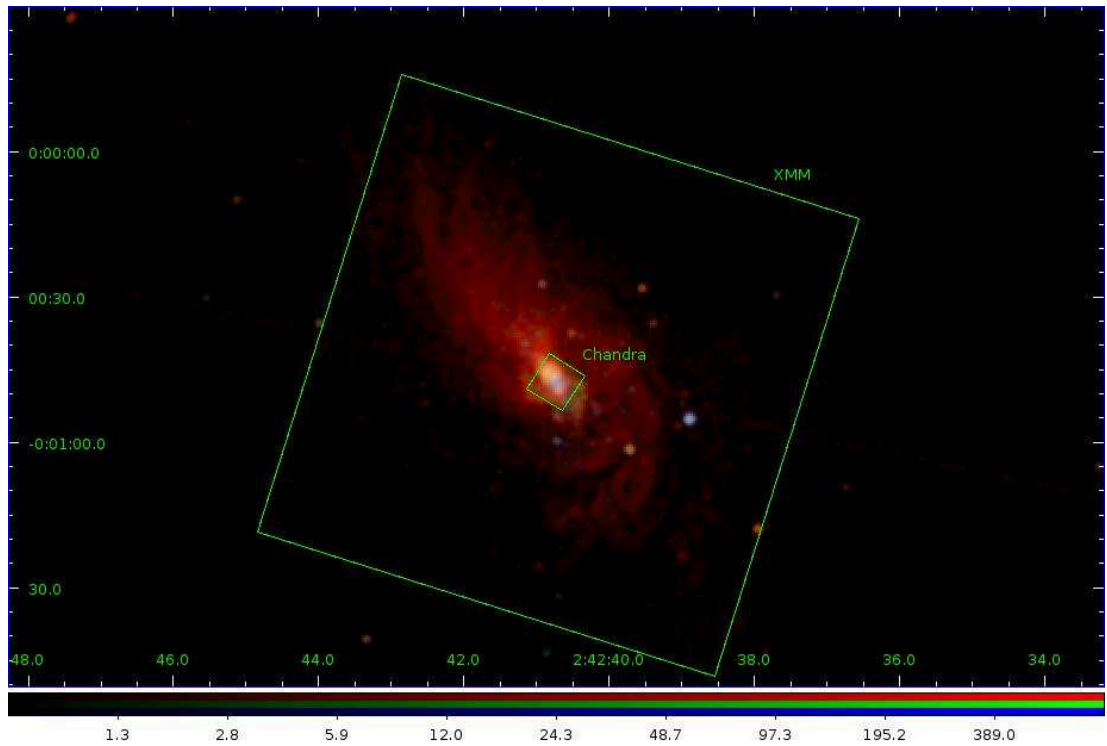}  

\caption{XMM/RGS and {\it Chandra} extraction cells, overlaid on the {\it
Chandra}/ACIS image.  Declination is along the y-axis and
Right Ascension is along the x-axis and the colors in the image correspond to the following:
red, 0.2 - 1.5 keV;  green, 1.5 - 2.5 keV; and blue, 2.5 - 8.0 keV.  The 
 extraction cells are square with size of $\sim 100^{''}$  and  $\sim 9^{''}$
for XMM/RGS  and {\it Chandra}/HETG, respectively, oriented with respect
to the roll angles for these observations \citep{kinkhabwala2002, kallman2014}.}  

\label{Windows}
\end{figure}

As expected, our measured emission-line fluxes are in good agreement with
those of \citet{kinkhabwala2002}. However, the RGS fluxes are $\sim$ 2 greater
than those measured in {\it Chandra}/HETG spectra \citep{ogle2003,
kallman2014}. The difference is due to the much larger extraction region for the
RGS. To illustrate this, in  Figure~\ref{Windows} we show the RGS and 
{\it Chandra}/HETG extraction windows, overlaid 
on a {\it Chandra}/ACIS image. Clearly, much of the X-ray emission-line region is 
outside the {\it Chandra} window.

Fe M- and L-shell transitions (Fe {\sc xiv} to Fe {\sc xxiv})  are heavily blended with H- and He - like lines of Ne (Figure~\ref{whole spectra}), therefore the strengths of these lines could not be usefully constrained. 
The higher-order transition lines of Mg, Si and S are barely resolved  due to the low sensitivity of the RGS in the wavelength regime (6-10 {\AA}).

\subsubsection{Kinematics}
\label{Kinematics}

\begin{figure}[ht]

\includegraphics[scale=0.66,viewport=5 0 500 540]{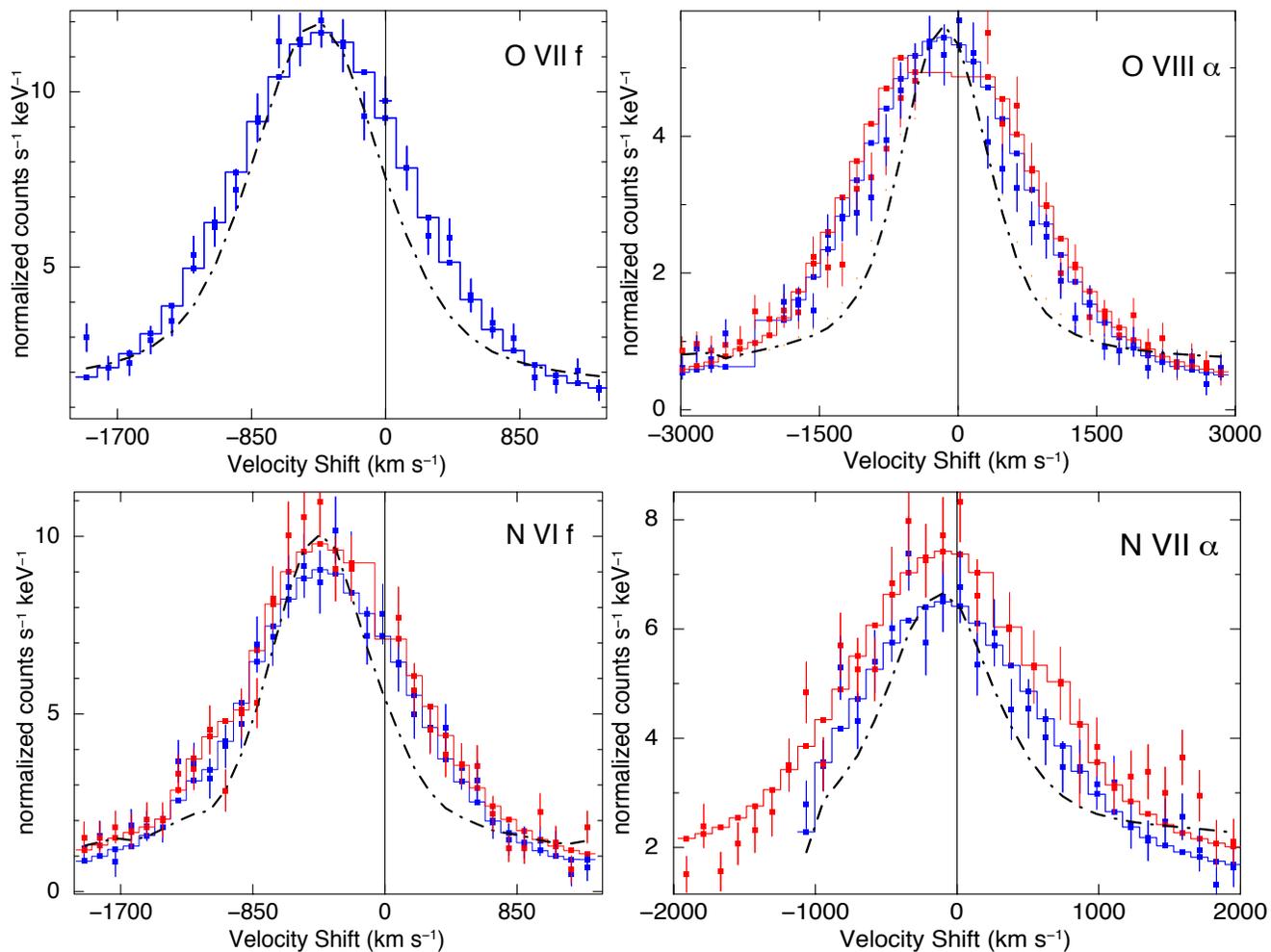}  

\caption{ RGS 1 ({\it red}) and RGS  2 ({\it blue}) line profiles for O {\sc vii} {\it f}, O {\sc viii} Ly$\alpha$,  N {\sc vi} ${\it f}$ and N {\sc vii}, relative to the nucleus. 
The systemic velocity of the host galaxy is represented here by the solid line at 0 km s${^{-1}}$.  
Note the blueshifts of each emission-line peak. Each line is broaded than the instrument profile which is represented
by the dotted line. }  

\label{velocity profiles}
\end{figure}

We have identified the observed lines using 
 the expected laboratory energies from the National Institute of Standards and Technology (NIST), supplemented by  
  the Kentucky atomic database (Table~\ref{velocity shift}). 
 Most of the emission lines show a significant blue-shift relative to the host galaxy \citep[cz = 1137 ${\pm}$ 3 km/s;][]{huchra1999}, 
 indicating an origin in outflowing 
 gas, as previously found by other authors \citep[e.g.][]{kinkhabwala2002}. 
 Based on the strongest, most isolated lines from He-like N and O and H-like
 C, N, and O, there is evidence for two kinematic components, with the 
 N~{\sc vi} and O~{\sc vii} {\it f} lines having more negative radial velocities than
 the N~{\sc vii} and O~{\sc viii} Ly$\alpha$ lines (See Figure~\ref{velocity profiles}). However, the velocities
 of the latter are consistent with those of the N~{\sc vi} and O~{\sc vii} {\it r}
 lines, within the uncertanties. Unfortunately, the Ne~{\sc ix} and Ne~{\sc x}
 lines are weaker and heavily blended with Fe lines, hence it is impossible
 to see if the same trend is present. Overall, the radial velocities are on the
 same order as that measured for [O~III] $\lambda$5007, 
 $v_{\rm rad}\sim$ 160 km s$^{-1}$ \citep{crenshaw2010}, which suggests that the
 X-ray and optical emission-line gas have similar kinematics. Note that these
 velocities are not de-projected and the actual outflows may be much
 faster \citep[e.g.][]{vdas2006}. As we also note in Section~\ref{Final Fit}, 
 while the kinematics are {\it consistent} with distinct
 low and high ionization zones, their radial velocities cannot be well
 constrained with these data. 

Velocity widths ($FWHM$) were determinable for a few strong lines 
(Table~\ref{velocity shift}).  The strongest lines, e.g., 
N {\sc vi} ${\it f}$ and O {\sc vii} ${\it f}$ are 
resolved using RGS, with velocity 
widths $FWHM \approx 926$ and ${\approx}$ 1003 km s$^{-1}$, respectively\footnote{The
RGS FWHMs at these wavelengths, linearly extrapolating between the values at 35.4\AA~ and
15.5\AA~, are 775 km s$^{-1}$ and 1000 km s$^{-1}$}.
This is quite
close to the value measured for [O~{\sc iii}] $\lambda$5007, $FWHM \approx 1060$ km s$^{-1}$ \citep{whittle1992},
again indicative of similar kinematics for the X-ray and optical emission line gas.
The values are in
excess of the radial velocities and most likely result from the superposition
of different kinematic components along our line-of-sight.

\section{Photoionization Modeling} 
\label{Photoionization Modeling}

\subsection{Inputs to the Models}
\label{Model Inputs}

Previous analyses for the soft X-ray emitting gas in NGC~1068 have used selected line measurements to explore the conditions in the emitting gas \citep{kinkhabwala2002}. 
Here we aim to construct a self-consistent photoionization model for the X-ray emitter.  To this end,  we have made use of the photoionization code, {\sc cloudy} version C10.00, 
last described by \citet{ferland2013}. As usual, our model results depend on the choice of input parameters, specifically:
the spectral shape of the incident radiation or spectral energy distribution (SED), the radial distances of the emission-line gas with respect to the
central source, the number density of atomic hydrogen ($n_{\rm H}$) and column density ($N_{\rm H}$) of the gas, and its chemical composition. Given the large radial distances
($>$ 10s of pc) of the emission-line gas, we have assumed open, or slab-like, geometry. The models are parameterized in terms
of the dimensionless ionization parameter $U$, the ratio of ionizing photons per nucleon at the illuminated face of the slab, or:
 
\begin{equation} 
 U = \frac{Q}{4\pi~ R^2~ c~ n_{\rm H}}
\end{equation} 
 
where, $R$ is the distance to the continuum source, c is the speed of light, and the
total number of Lyman-continuum photons sec$^{-1}$ $Q =\int_{13.6 eV}^{\infty}\frac{L_{\nu}}{h\nu}~d\nu$,
emitted by a source of luminosity $L_{\nu}$.

\subsubsection{The Ionizing Continuum}
\label{Ionizing Continuum}

We assumed an SED in the form of a broken power law F${_\nu}$ = K ${\nu^{-\alpha}}$, with 
       ${\alpha}$ = 1.0 below 13.6 eV,  ${\alpha}$ = 1.7 from 13.6 eV to 0.5 keV, and ${\alpha}$ = 0.8 from 0.5 keV to 30 keV 
       \citep[][]{kraemer&crenshaw2000}, where $\alpha$ is the energy index. We also included a low energy cut-off at 1 eV and a high energy cutoff
at 100 keV.

Since our view of the central source is blocked in NGC~1068, we estimated the ionizing luminosity using
an ``isotropic'' quantity, specifically the [O~{\sc iv}] 25.89$\mu$m emission line (see \citet{melendez2008}). The [O~{\sc iv}] flux, detected with the {\it Infrared Space Observatory}-Short Wave Spectrometer, is approximately 1.9 $\times$ 10$^{-11}$ ergs~cm$^{-2}$~s$^{-1}$ \citep{lutz2000}, which corresponds to
a line luminosity $L_{\rm OIV}$ $\approx$ 10$^{41.53}$ ergs s$^{-1}$. Using the
linear regression fit to $L_{\rm OIV}$ and the 2-10 keV (L$_{2-10 keV}$)
luminosity for Seyfert 1 galaxies calculated by 
\citet{melendez2008}, we estimate $L_{2-10 \rm keV} \sim$ 10$^{43.8}$ ergs s$^{-1}$, which, for a bolometric
correction of $\sim$ 30 \citep{awaki2001}, yields $L_{\rm bol} \sim$ 10$^{45.3}$ ergs s$^{-1}$. The corresponding 
mass accretion rate is $\dot{M} = L_{\rm bol}/\eta c^{2}$, or 0.35 M$_\odot$ yr$^{-1}$, for $\eta = 0.1$. Interestingly, given
a black hole mass of 1.5 $\times$ 10$^{7}$ M$_\odot$ \citep{greenhill1997}, the central source is radiating at approximately its Eddington
limit. Based on these values and our assumed SED, we find the ionizing luminosity $L_{ion} \sim 10^{44.2}$
ergs s$^{-1}$ and $Q \sim 10^{54.4}$ photons s$^{-1}$. 

\subsubsection{Elemental Abundances}
\label{Abundances}

Our initial approach was to find an approximate solution for the gas by comparing the ratio of intensities of selected strong emission lines in the data with 
those predicted by the CLOUDY model. The line ratios selected were for the strengths of H-like and He-like ions relative to O {\sc vii} {\it f} ${\lambda}$22.10\AA~ 
(the latter was selected by virtue of being the strongest line detected in the RGS data), similar to the approach used in the analysis of the
RGS spectrum of the Seyfert 1 galaxy NGC~4151 by \citet{armentrout2007}.  
We first generated {\sc cloudy} photoionization models assuming solar abundances \citep{grevesse1998}.
This initial comparison revealed no single-zone gas solution that could adequately describe the observed line ratios. Specifically, models with 
solar abundances over-predicted the strongest oxygen lines, O{\sc vii} {\it f} ${\lambda}$22.10\AA~ and O {\sc viii} Ly$\alpha$, relative to the lines from other abundant elements/ions, in particular,  and He-like carbon and nitrogen. Interestingly, in photoionized gas
the X-ray emission lines from second-row elements are primarily produced by recombination or, for permitted transitions 
if the lines are optically thin, by photoexcitation. The other strong features in the RGS spectrum include radiative recombination
continua (RRCs) which are obviously formed via recombination. Hence, the strengths of these features are quite sensistive to relative
elemental abundances. Notably, it has been argued that optical and UV recombination lines are more reliable
indicators of elemental abudances than collisionally excited forbidden lines, which are typically used to estimate 
elemental abundances in optical spectra, due to the sensitivity of the latter on temperature and density fluctuations
\citep{peimbert1967}.

In the analysis of RGS spectra of the Seyfert 1 NGC 3516,
\citet{turner2003} found that the model fit was improved by assuming both
super-solar N/O and sub-solar C/O abundance ratios. 
They argued that this was consistent with conversion of carbon to nitrogen via 
the CNO-cycle in intermediate mass stars (M$ \leq$ 7 M$_{\sun}$; e.g. \citet{maeder1989}),
and that fairly large N/O ratios, e.g., a few times solar, could
occur in stars with initially roughly solar abundances. However, the NGC~1068 spectra
also show strong carbon lines (see Fig. 1), which are not consistent with the
loss of carbon, at least if the overall abundances were approximately solar. In studies of H~II regions in the Milky Way, evidence
has been found for a C/O gradient, with a C/O ratio $\sim$ unity, while O/H ratio was roughly
solar, at radial distances
$\lesssim$ 7 kpc \citep[e.g.][]{esteban2005}. The enhancement of carbon in the Milky
Way interstellar medium could be due to either stellar winds from massive stars,
8 $\leq$ M/M$_{\sun}$ $\leq$ 80
\citep{henry2000}, or a combination of high-metallicity massive stars and low-metallicity
low-to-intermediate mass stars, 0.8 $\leq$ M/M$_{\sun} \leq $ 8 \citep{carigi2005}. In either case,
if the carbon were significantly enhanced in a star with otherwise solar abundances, the conversion
of carbon-to-nitrogen described above could lead to both high C/O and N/O ratios while O/H evolved
as $Z/Z_{\rm o}$.
Based on this, we assumed that metallicities of pro-genitor stars were solar \citep{asplund2005}, although with
C/O approximately 3 $\times$ solar, as compared to the roughly twice solar values found by \citet{esteban2005}.
We further assumed that nucleosynthesis brings the overall
metallicity to 1.5 $\times$ solar, but with all the added carbon going into the production of nitrogen.  The resulting
abundances,  
with the log values, relative to H by number, are as follows:
He: $-$0.83; C: $-$3.06;  N: $-$3.36;  O: $-$3.13;
Ne: $-$3.90; Na: $-$5.59; Mg: $-$4.23; Al: $-$5.39;                                              
Si: $-$4.32; P: $-$6.42;  S: $-$ 4.71; Ar $-$ 5.43;                                          
Ca: $-$5.49; Fe $-$4.33; and Ni: $-$5.61.   
Here the N/H and C/H ratios are 6 and 3.2 $\times$ solar, respectively,  while the other heavy element abundance ratios are 1.5 $\times$ solar.
It should be noted that \citet{brinkman2002} suggested super-solar nitrogen abundance and \citet{kallman2014} required non-solar
heavy element abundances for their photoionization models.
Based on the strong C and Fe emission in the RGS spectra, there is no evidence for depletion of
these elements onto dust grains. Therefore we did not include cosmic dust in the models.

\subsection{Spectral fitting results}
 \label{Atablefit}

The range in ionization states detected in these spectra suggests the presence of
two distinct components of emission-line gas, therefore our approach was to generate two
model grids with {\sc cloudy}. Consideration of the ratio 
$\frac{\rm N {\rm VI} {\it f}} {\rm O {\rm VII} {\it f}} $ and 
${\frac{\rm Ne {\rm X}} {\rm O {\rm VIII}}}$ suggested that 
these zones lie in the ranges log$U \sim -1$ to  0, log$N_{\rm H} \sim 20 - 22$ and 
log$U \sim 0 - 2$, log${N_{\rm H} \sim} 20 - 23$.

\subsubsection{The Final Model}
\label{Final Fit}

To refine our solution, the RGS spectra were compared to a model comprising  two `candidate' model zones.  Based upon the initial results from the line ratio analysis, 
 we re-ran {\sc cloudy} with model step intervals of 0.1 in the log of $U$ and $N_{\rm H}$, across the ranges $ -2 <  {\rm log}U <  2$ and $ 20 < {\rm log}N_{\rm H} < 24$,
 using the elemental abundances noted above. Initially, the resonance lines of the He-like triplets (only) were underpredicted compared to the forbidden lines. 
 Thus, to boost the strength of these resonance lines, we included 
micro-turbulence of 35 km s${^{-1}}$. This corresponds to a {\it FWHM} $=$ 82 km s$^{-1}$, which is significantly
less than that of the resolved lines discussed in Section~\ref{Kinematics}, which likely result from the
superposition of kinematic components along our line of sight.
Following \citet{porter2006}, we then created a FITS format {\sc atable} from the {\sc cloudy} output. 
To facilitate a comparison of the model tables with the ensemble of line results we performed spectral fitting of the RGS data using our {\sc atable}. 

Our final model was comprised of two zones (LOWION and HIGHION), each absorbed by the Galactic line-of-sight column density,
 parameterized using {\sc tbabs}. The value of {\sc tbabs} was initially set to 
 $N_{\rm H} = 3.53 \times 10^{20}  {\rm cm^{-2}}$\citep{kalberla2005}, but in the final fit the 
value was allowed to vary.
Fitting the model to the data showed the model lines to be 
systematically too narrow to account for the observed line ensemble. 
Therefore we smoothed the model spectra using a Gaussian function equivalent to $\sigma = 365$ km s$^{-1}$ ($FWHM = $852 km s$^{-1}$)
for LOWION and $\sigma = 1170$ km s$^{-1}$ ($FWHM = 2732$ km s$^{-1}$) for HIGHION,
The LOWION smoothing is consistent with the measured widths for
N {\sc vi} ${\it f}$ and O {\sc vii} ${\it f}$ (see above). 
The fitting was allowed to proceed until the reduced $\chi^{2}$ was minimized; the
final model parameters are listed in Table~\ref{model parameters}. 
In the process, the outflow velocities for LOWION and HIGHION converged on values
of 215 km s$^{-1}$ and 166 km s$^{-1}$, respectively. However,
these values overlap within the uncertainties, as suggested in Section~\ref{Kinematics}.
Also, the fitting returned a Galactic column of 5.43$^{+0.44}_{-0.28}$ $\times
10^{20}$ cm$^{-2}$, somewhat larger than the value from \citet{kalberla2005}. However,
there is evidence for absorption by low-ionization gas covering the NLR of NGC
1068 (see \citealt{kraemer&crenshaw2000}; \citealt{kraemer2011}), and it is plausible
that the extra column of neutral gas is associated with that.

\begin{figure}[ht]

\includegraphics[scale=0.85,viewport=120 0 600 500]{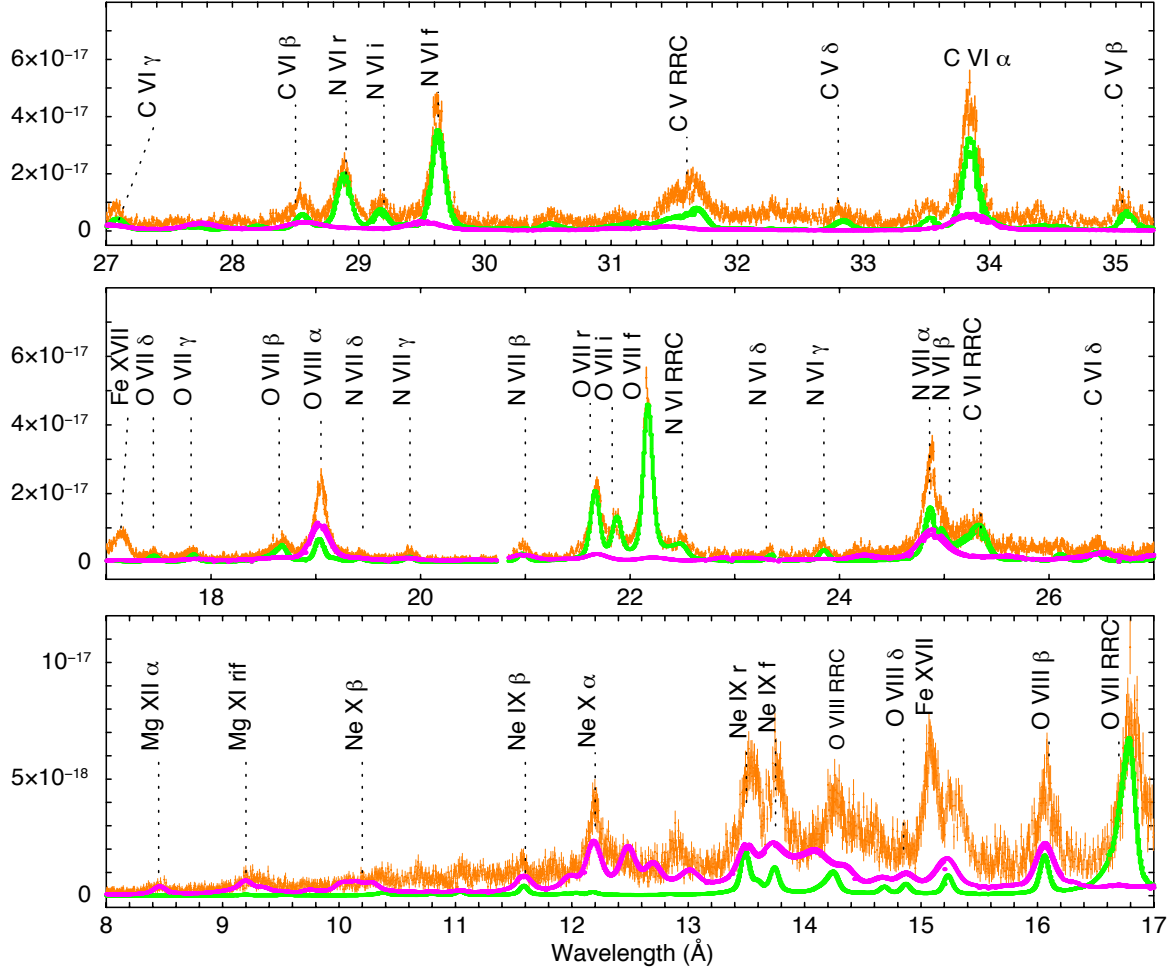}  
\vspace{-0.35in}
\caption{The individual model components, LOWION (green) and HIGHION (purple) compared to the combined RGS1 and RGS2 spectrum.
As noted in the text, LOWION contributes most of the emission from the He-like C, N, and O, while HIGHION contributes most of the H-like N and O 
and both H and He-like Ne emission.}
 
\label{components spectra}

\end{figure}

\begin{figure}[ht]

\includegraphics[scale=0.85,viewport=120 0 600 500]{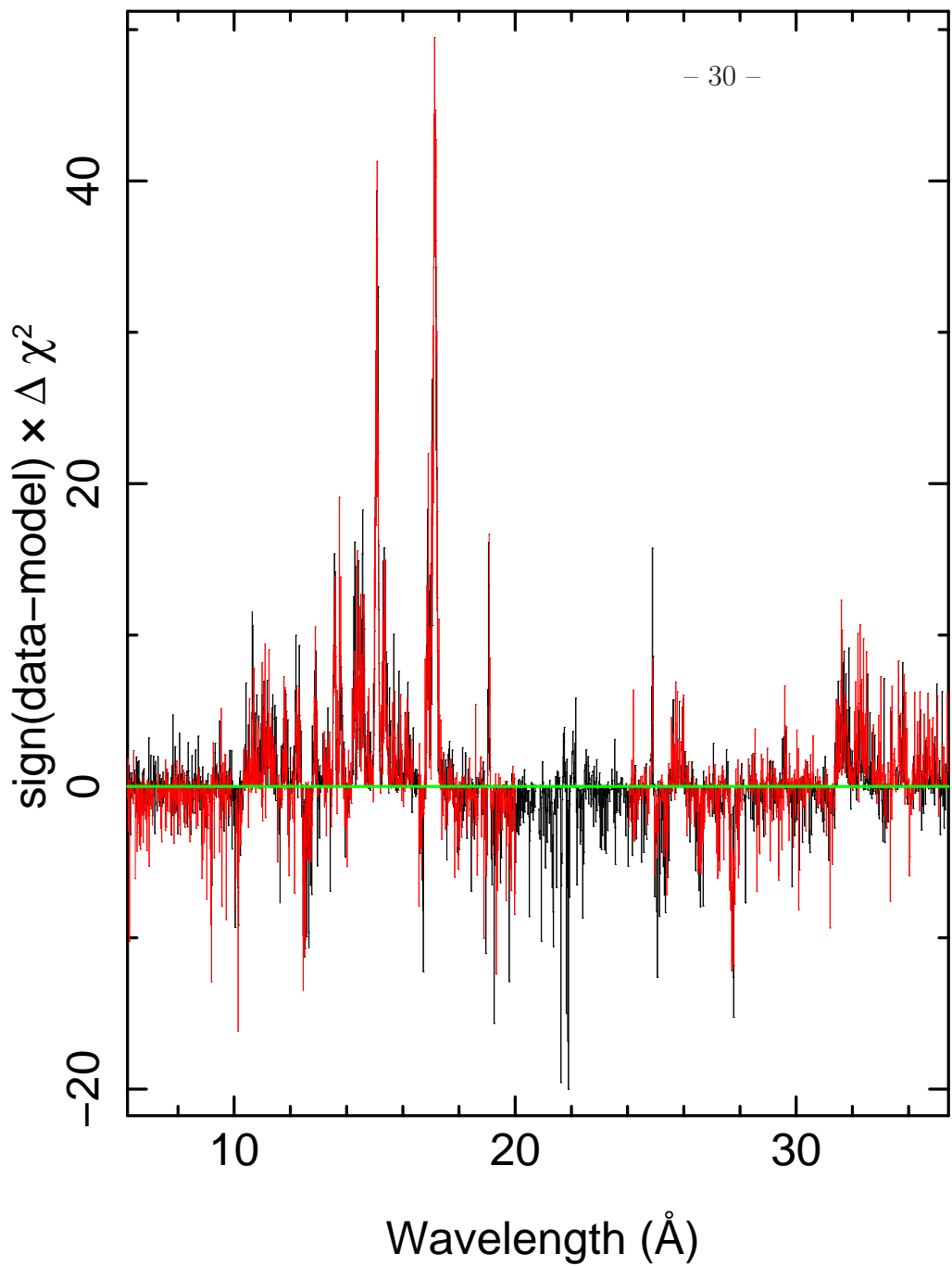}  
\vspace{-0.5in}
\caption{The contributions to $\chi^{2}$ for the data compared to 
the best fitting two-zone model from {\sc xspec}, detailed in 
Section~\ref{Photoionization Modeling}. RGS 1 data are shown in black, RGS 2 data are shown in red. 
The gaps are where the dispersed RGS spectrum falls upon a non-operational 
CCD chip. The greatest mis-match occurs in the range of 14\AA--17\AA~ and
is primarily due to the under-prediction of emission lines from M-shell
Fe ions.}

\label{chi2 fit}

\end{figure}

As shown in Figure~\ref{whole spectra}, we obtained a reasonable (i.e., reduced $\chi^2$
$=$ 1.99) fit to the data with two components, the relative contributions of which are
shown in Figure~\ref{components spectra}. The contributions to $\chi^2$ are shown in
Figure~\ref{chi2 fit}, which illustrates that most of the
mismatch to the data is due to under-predicted emission below
17\AA.  In order to 
determine and compare the predicted emission line fluxes to the
intrinsic line luminosity, which is necessary to determine the covering
factors of the emission-line gas, $n_{\rm H}$ must be
determined for each
component, which requires fixing their radial distances.
For the sake of simplicity, we assumed that LOWION and HIGHION 
are co-located. As a reference point, we assume they are at a distance $R = 50$ pc
from the central source, which is consistent with the fact that most of the X-ray emission
arises within the central $\sim$ 100 pc \citep[see][]{ogle2003}. Note that the X-ray emission-line ratios are not 
sensitive to density over the range expected for X-ray emitters in the NLR (e.g., $<$
10$^{6}$ cm$^{-3}$; \citealt{porquet&dubau2000}). Also, we do not have any strong
constraints of the location of the emission-line gas, except that the emission
is centrally peaked. However, this distance is reasonable given the distribution of
scattered continuum \citep{crenshaw&kraemer2000A} and the probability that much of
the emission from the inner 30 pc is heavily attenuated (e.g., 
\citealt{kraemer&crenshaw2000C}; \citealt{kraemer2011}). Using the value of $Q$ derived in
Section~\ref{Ionizing Continuum}, we obtain $n_{\rm H}$ = 15 cm$^{-3}$ and 265 cm$^{-3}$ for HIGHION and
LOWION, respectively. One additional check on whether the models
 are physically plausible is the
requirement that the depth of the model, $\Delta R = N_{\rm H}/n_{\rm H}$, is less
than the components radial distance, $R$ \citep[e.g.][]{blustin2005, crenshaw2012}. For these model 
parameters, $\Delta R/R$ $=$ 0.02 and 0.68 for LOWION and HIGHION, respectively.

Based on the fitting, LOWION and HIGHION contribute 0.96 and 0.04 of O~{\sc vii}-f and
0.14 and 0.86 of O~{\sc viii} Ly-$\alpha$, respectively. Using these fractional contributions, we
computed the predicted emission-line fluxes from each component and the total
flux. The former are computed by comparing the predicted O~{\sc vii}-f and O~{\sc viii} Ly-$\alpha$ fluxes
to the absorption-corrected fluxes, taking into account the fractional contributions of each
component, in order to derive a scaling factor for each component. Then the 
remaining line fluxes are computed by multiplying the ratios of their fluxes to those of 
O~{\sc vii}-f or O~{\sc viii} Ly-$\alpha$ by the derived scaling factors.
The final values are listed in Table~\ref{observations}, along with the observed and absorption-corrected 
fluxes.  The fits for the
individual lines are good overall, with the predictions for most of the stronger lines (i.e.,
with fluxes $\gtrsim$ 3 $\times$ 10$^{-4}$ photon cm$^{-2}$ s$^{-1}$) within
$\sim$ 30\% of the absorption-corrected values. The discrepancies include
C~{\sc v} He$\beta$ and He$\delta$, which are relatively weak and in a region where
determining the continuum level is non-trivial and N~{\sc vii} Ly-$\alpha$ is somewhat
under-predicted.

In addition to emission lines included in Table~\ref{observations}, we also
compared the
predicted and measured RRCs, using the same relative contributions from the two model
components (see Table~\ref{rrcs}). Although we detected the O~{\sc VIII} RRC, the feature is too heavily blended
with the surrounding emission for us to have been able to determine the width and flux 
accurately. Overall, the predicted fluxes fit the measured values reasonably
well. The model-predicted electron temperatures correspond to widths of k$T$ = 4.0 eV and 44.4 eV, for
LOWION and HIGHION, repsectively. While the LOWION value is on the same order as the
measured values of the He-like RRCs, the HIGHION prediction is several times that of the 
measured H-like RRCs. This is likely the result of the difficulty in fitting these
broader features. However, both the models and data confirm that the emission-line gas
possesses temperatures consistent with photoionization.

As noted above, the greatest mismatch between the model and the data is
in the region 10-17 \AA~, which shows a heavy blend of emission
lines of Fe {\sc xiv}-{\sc xxiv} with those from Ne~{\sc ix}, Ne~{\sc x}, and O~{\sc viii} 
(for a more complete identification of the iron lines see \citealt{kinkhabwala2002}).
HIGHION predicts that the maximum ionization states of iron are spread from
Fe~{\sc xvii} to Fe~{\sc xxi}, within the observed range, however the predicted line fluxes
are quite weak and essentially make a negligible contribution to the model
spectrum (see Figure~\ref{components spectra}). The underprediction may be the result 
of the incomplete atomic data for iron, specifically rates for fluorescence following
inner shell ionization.

\subsubsection{Model-derived Covering Factors}
\label{Covering Factors}

Having obtained a reasonable fit to the RGS spectrum with our two component
model, we calculate the covering factors ($C_{f}$) for each component by
comparing the emitting area, the ratio of the emission-line luminosities to
their model-predicted fluxes, to the surface area of a sphere surrounding the
central source. As we mentioned above, we assume that both LOWION and HIGHION 
are 50 pc from the source, which sets $n_{\rm H}$ for each component for the value
of $U$ for each component, and thus the predicted emission line-fluxes. Based on
our spectral fitting, the total luminosity of the O~{\sc vii}-f line emitted by
LOWION is $\sim$ 2.5 $\times$ 10$^{40}$ ergs s$^{-1}$. Dividing by the 
predicted flux, 0.215 ergs~cm$^{-2}$~s$^{-1}$, the total emitting surface area of LOWION is $\sim$ 10$^{41}$
cm$^{2}$, which corresponds to a covering factor $C_{f} = 0.35$. Based again on
our spectral fitting, for HIGHION,
the total luminosity of O~VIII Ly$\alpha$ is $\sim$ 1.5 $\times$ 10$^{40}$
ergs~s$^{-1}$, while the predicted flux is 0.051 ergs~cm$^{-2}$~s$^{-1}$, from which we derive an emitting surface of 2.5 $\times$
10$^{41}$ cm$^{2}$ and $C_{f} = 0.84$. Although the covering factors are physically possible,
in the sense that they are less than unity, the value for HIGHION
requires that this component subtends a larger solid angle than the
emission-line bicone, even if the bicone were filled \citep[e.g.][]{vdas2006}. 
We will revisit this point in Section~\ref{structure}.

\subsubsection{UV and Optical Contraints on the X-ray Emission-Line Gas}
\label{UV Constraints}

Although the ionization parameters for both emission components are
significantly higher than those determined for the UV and optical emission-line gas
\citep[e.g.][]{krc1998, kraemer&crenshaw2000}, except for the ``CORONAL''
component of the Hot Spot \citep{kraemer&crenshaw2000C}, the LOWION model
predicts strong O~{\sc vi}~$\lambda\lambda$ 1031.9,1037.6 
Scaling the predicted flux by the emitting area determined
from the O~{\sc vii} f-line, the predicted O~{\sc vi} luminosity is $\sim$ 1.9 $\times$ 10$^{41}$ ergs
s$^{-1}$, which corresponds to an observed flux F$_{OVI}$ $\sim$ 1.0 $\times$ 10$^{-11}$
ergs cm$^{-2}$ s$^{-1}$ (note that the contribution from HIGHION is more than
two orders of magnitude less, hence can be ignored).

NGC 1068 was observed with the Hopkins UltraViolet Telescope (HUT), aboard the space
shuttle Columbia \citep{kriss1992}, through two circular apertures of 18$''$ and
30$''$, hence encompassing the region of strong X-ray emission
\citep[e.g.][]{young2001}. They
measured F$_{O VI}$ $=$ 3.74 ($\pm$ 3.1) $\times$ 10$^{-12}$ ergs cm$^{-2}$
s$^{-1}$. While the emission-line fluxes reported by Kriss et al. were not corrected for
extinction, based on the the ratio of He~{\sc ii}~$\lambda$1640 (from HUT) to
He~{\sc ii}~$\lambda$4686 \citep{koski1978}, they derived an extinction {\it E$_{B-V}$} $=$
0.16, assuming an intrinsic 1640/4686 ratio of 7.0 \citep{seaton1978}. Using the UV
extinction curve in \citet{seaton1979}, the corrected F$_{O VI}$ $\approx$ 2.5 $\times$ 10$^{-11}$
ergs cm$^{-2}$ s$^{-1}$, which indicates that LOWION contributes 40\% of the
O~{\sc vi}
emission. Based on this, our model meets the constraints from the UV emission.
Also, it is possible that some regions of UV and optical emission are
undetectable due to extinction (see discussion in \citealt{kraemer2011}), but
could be detected in the X-ray, in which case LOWION accounts for an even smaller
fraction of the intrinsic UV emission.

Given the model parameters of the Hot Spot CORONAL component (Log$U = 0.23$;
log$N_{\rm H} = 22.6$; \citealt{kraemer&crenshaw2000C}), it
could contribute to the overall X-ray emission. However, neither of our components
have similar parameters. Furthermore, while CORONAL was optimized to match the observed
[S~{\sc xii}] $\lambda$7611, the peak ionization states are S~{\sc ix} and S~{\sc xv} for LOWION and
HIGHION, respectively. Finally, forcing the inclusion of a component similar to CORONAL
into the spectral fitting produced statistically unacceptable results. This suggests 
that the conditions that give rise to the [S~{\sc xii}] emission are not typical of the X-ray
NLR as a whole.

\section{Discussion}
\label{Discussion}

\subsection{Total Mass and Mass Loss Rates}
\label{Mass Loss}

Based on our model results and constraints on the covering factors of each component, we
can determine the total mass of the X-ray emission-line gas, $M_{\rm tot}$, for a given radial
distance, $R$. Assuming the gas lies in shells, the thickness of which are constrained by the model-derived
values of $N_{\rm H}$, the total mass is given by:

\begin{equation}
M_{tot} = 4\pi R^{2} N_{\rm H} \mu {\rm m_{\rm p}} C_{f}
\end{equation}  
where m$_{\rm p}$ is the mass of a proton and the factor $\mu$ is the mean
atomic mass per proton (we assume $=$ 1.4, primarily from the contribution
from helium). For our fiducial distance, $R = 50$pc, and derived model parameters,
we obtain $M_{\rm tot} = 8.7 \times 10^{4}$ M$_\odot$ and $4.7 \times 10^{5}$ M$_\odot$ for LOWION and
HIGHION, respectively. Note, given the possibility that the emission-line gas lies at
greater radial distance, $M_{\rm tot}$ could be somewhat greater. 

We estimate the mass loss rates, $\dot{M}$, as follows \citep{crenshaw2003}:

\begin{equation}
\dot{M} = 4 \pi R N_{\rm H} \mu {\rm m_{\rm p}} C_{f} v 
\end{equation}
where $v$ is the outflow velocity. Using $v_{\rm rad}$ in place of $v$, we obtain
mass loss rates of $\dot{M} \sim 0.38$ M$_\odot$ yr$^{-1}$ and
0.06 M$_\odot$ yr$^{-1}$ for the two components. Clearly, $\dot{M}$ would be greater
if the gas were at a larger radial distance, as is the case for $M_{\rm tot}$, or if the ouflow velocities were
greater than $v_{\rm r}$. 

The derived values of $M_{tot}$ and $\dot{M}$ are roughly the same as those determined
from the {\it Chandra}/HETG spectrum by \citet{kallman2014}, and, based on
our estimate of $L_{\rm bol}$, the latter is on the same order as that of the fueling rate. From the results of the
STIS long slit spectral analysis \citep[]{kraemer&crenshaw2000C, kraemer&crenshaw2000}, we estimate
that within a single slit position, at PA202\deg, the total amount of emission-line gas is $\sim  6 \times 10^{3}$ M$_\odot$.
Scaling this quantity by the ratio of the dereddened [O~{\sc iii}] $\lambda$5007 flux from the STIS
spectra and the [O~{\sc iii}] flux in ground-based spectra \citep{bonatto1997}, dereddened as per the discussion in 
Section~\ref{UV Constraints}, we estimate a total
mass of the optical emission-line gas of $\sim$ 3.8 $\times$ 10$^{4}$ M$_\odot$. This is
on the same order mass of ionized gas determined from spectra obtained with the 
{\it Gemini}/Near-Infared Field Spectrograph by \citet{riffel2014}, hence it is unlikely that we have
grossly underestimated the mass of optical emission-line, due to regions of heavy extinction 
\citep[e.g.][]{kraemer2011}. Therefore, these results indicate that the X-ray emission-line gas is
a major, if not dominant, component of ionized gas in the NLR.

\subsection{Thermal and Pressure Stability}
\label{Stability}

The predicted gas pressures at the ionized faces of LOWION and HIGHION are 4.3 $\times 10^{-9}$ dyn~cm$^{-2}$
and 2.7 $\times 10^{-9}$ dyn 
~cm$^{-2}$, respectively, which indicate that, given uncertainties in the model parameters, these components are roughly in
pressure equilibrium if co-located. In contrast, the predicted gas pressure for the UV/optical emission-line gas from the Hot Spot is
3.5 $\times 10^{-7}$ dyn~cm$^{-2}$ (based on the model parameters in \citealt{kraemer&crenshaw2000C}). Therefore,
UV/optical knots would not be pressure-confined by the X-ray emission-line gas. In that case, one would expect
to see a drop in the density of the optical/UV gas with radial distance. However, given the large pressure differential,
the density drop would
be much more rapid than observed \citep{kraemer&crenshaw2000}. This
suggests other scenarios, such as creation/evaporation of clouds out of/into the X-ray medium \citep{krolik2001} or
in situ acceleration of gas that has rotated into the illumination cone \citep{crenshaw2010B}, rather than outflow and expansion
of individual knots. 

\begin{figure}[ht]

\includegraphics[scale=0.76,viewport=85 0 500 540]{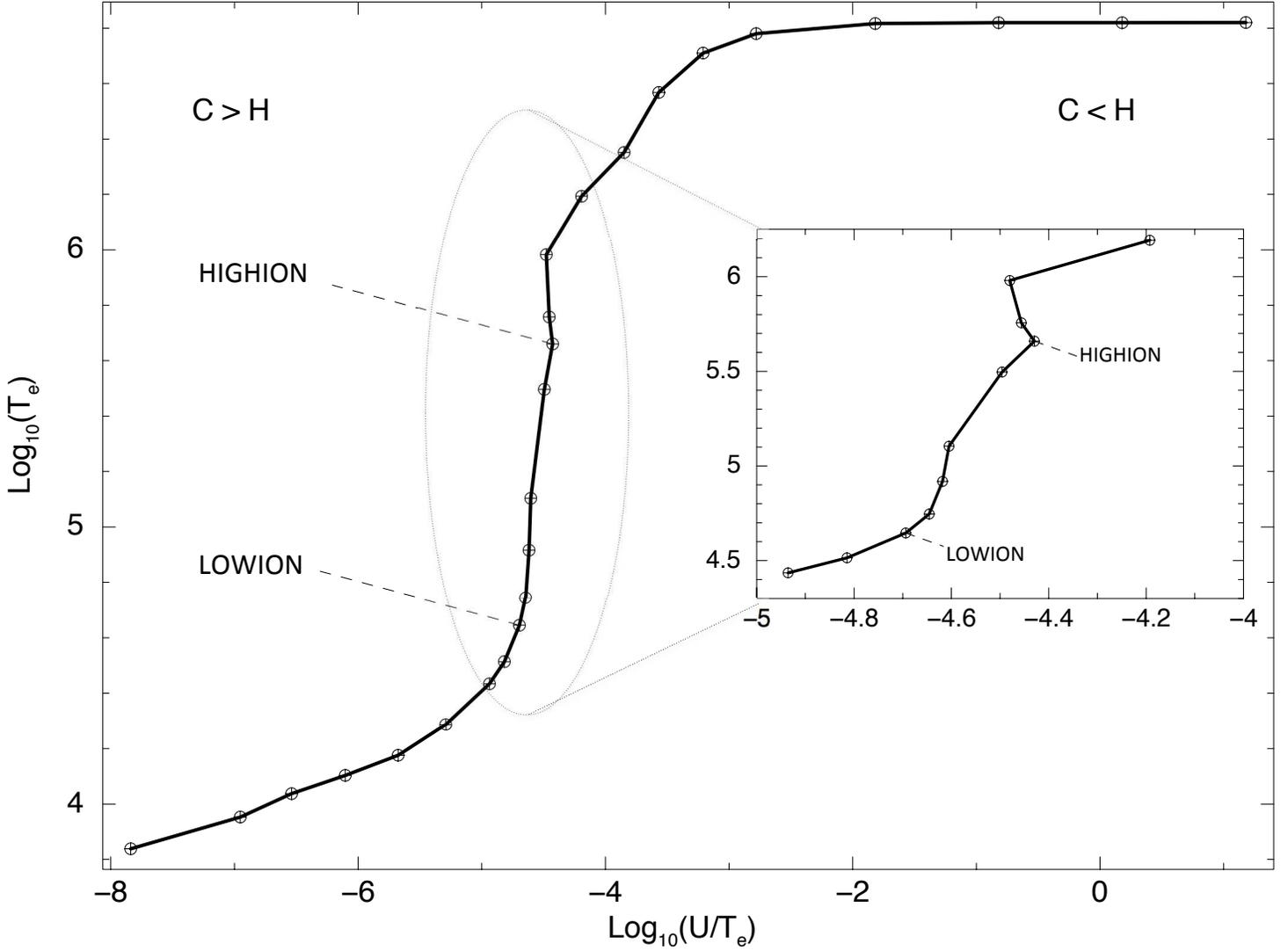}  

\caption{Stability curve for a range of ionization parameter from 10${^{-3}}$ to 10${^{7}}$ and our assumed SED
and abundances (see Section~\ref{Model Inputs}). The regions at the left and right of the curve are where the cooling and heating dominate, respectively,
as indicated. The two gas components identified from our photoionization modeling, LOWION and HIGHION, lie on stable, i.e., positively-sloped
portions of the curve (which is most clearly seen via the insert), although the latter is close to an unstable region.}  

\label{S_curve}
\end{figure}

In Figure~\ref{S_curve} we show the log$T -$ log$U/T$, or S-Curve, plot generated
with our assumed SED and abundances. Note that there is only one region with 
pronounced negative slopes, indicative of strong instabilities; the overall
stability is the 
result of the high metal abundances in our models \citep[see][]{bottorff2000}. The
insert shows that both components lie on stable, i.e., positive-sloped 
sections of the S-curve. While HIGHION does lie close to an unstable region,
given that the AGN is radiating close to its Eddington limit, it is unlikely
that it will experience an ionizing flux increase that would drive it into
instability.

\subsection{Structure of the X-ray NLR}
\label{structure}
    
As noted in Section~\ref{Covering Factors}, the model-derived covering factors
 are physically possible. However, it is difficult to reconcile such large values
considering that the optical emission SW of
the nucleus is heavily attenuated by the disk of the host galaxy \citep{kraemer&crenshaw2000C}
and that
X-ray emission from the inner $\sim$ 30 pc appears to be absorbed by gas outside the
bicone \citep{kraemer2011}. Hence, it is likely that there is more X-ray emission line gas 
than that detected in the
RGS spectra. If so, the covering factors could be significantly greater and 
could easily exceed unity for HIGHION.
 
\begin{figure}[ht]

\includegraphics[angle=90,scale=0.76,viewport=200 0 500 540]{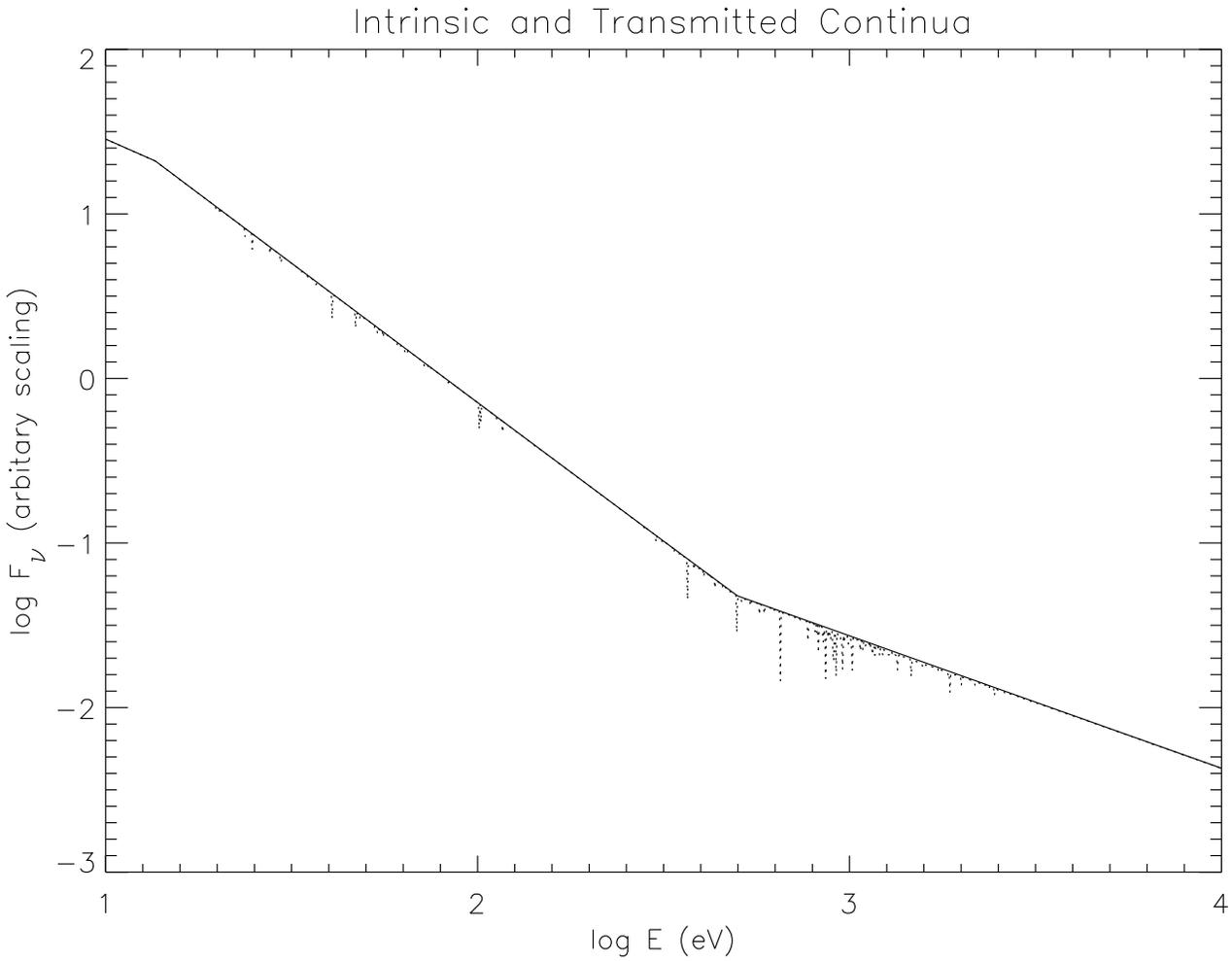}  
\vspace{2.0in}
\hspace{0.5in}
\caption{Incident (solid-line) and Transmitted (dotted-line) model continuua for the component HIGHION. As shown, there is little attenuation of the ionizing
radiation as it is transmitted, hence gas can be in a similar state of ionization in the ``shadow''
of the component.}  

\label{continua high}
\end{figure}

Both HIGHION and LOWION are matter-bounded and, in such cases, emission-line fluxes
can be increased to an extent by increasing $N_{\rm H}$. This would, correspondingly, decrease
the required emitting surface areas and, hence, the covering factors.  However,
in the case of HIGHION, $N_{\rm H}$ is 
constrained by the $\Delta$$R/R$ condition (see Section~\ref{Final Fit}).  
One way around this is if the emission-line gas consists of a number of matter-bounded components at
increasing radial distances. As an example,
in Figure~\ref{continua high}, we show the incident and transmitted continuum for HIGHION; clearly, there is no
significant attenuation of the incident continuum, hence highly-ionized gas could exist in the
``shadow'' of a component similar to HIGHION, i.e., subtending the same solid angle with 
respect to the ionizing radiation. LOWION also produces weak attentuation.
Assuming that density decreases with $R$, as observed with the optical emission-line 
gas \citep{kraemer&crenshaw2000C}, each additional zone could be at roughly the same ionization state. There
is some evidence for this scenario, in the sense that the optical 
continuum profile shows clear radial structure \citep{crenshaw&kraemer2000A}. A
series of separate shells would effectively create a large column density without
violating the $\Delta$$R/R$ constraint \footnote{Note that adding what are essentially identical
{\sc atables} would not change the fit to the RGS spectra, hence it is impossible
to test this scenario via {\sc XSPEC}.}, and thereby reduce the required covering factor.  
 
This proposed structure of the NLR has implications for the origin of the
polarized optical emission in NGC~1068, which is presumably due to scattering 
by free electrons.  
In their analysis of optical polarimetry of NGC~1068, \citet{miller1991}
determined that the temperature of the scattering medium is $\sim 3 \times$
10$^{5}$ K; for comparison, the predicted temperature for HIGHION is 5.15 $\times$ 10$^{5}$K.
However, the required column density of the medium is $\gtrsim$ 10$^{22}$ cm$^{-2}$ \citep{pier1994},
or an order of magnitude greater than that of HIGHION. On the other hand, if 
the emission line region is comprised of a series of zones, described above, the total
column density of high-ionization X-ray emission-line gas would be significantly larger than
that of HIGHION. Therefore, we suggest that gas with physical conditions similar to
those of HIGHION is the source of the scattered/polarized emission.

\section{Conclusions}
\label{Conclusions}

We analyzed an archival {\it XMM-Newton}/RGS spectrum of the Seyfert 2 galaxy
NGC~1068, which was previously published by \citet{kinkhabwala2002}. In the
process, we remeasured the emission-line fluxes, widths and radial velocities,
and the fluxes and widths of RRCs, and, overall, obtained similar values to
those of  \citet{kinkhabwala2002}. We generated photoionization models, using
Cloudy \citep{ferland2013},
to fit the emission-line spectrum. Our main results are as follows.

1. Overall the X-ray emission-lines have radial velocities blue-shifted with
respect to the systemic velocity of the host galaxy. The velocities and $FWHM$ are
roughly the same as that of [O~{\sc iii}] $\lambda$5007, which suggests that the X-ray
gas is part of the mass outflow through the NLR. Although there is evidence for
two kinematic components, with the 
 N~{\sc vi} and O~{\sc vii} lines having more negative radial velocities than
 the N~{\sc vii} and O~{\sc viii} lines, the differences are less
 than the measurement uncertainties.
 
2. Based on our preliminary modeling results, we determined  the 
abundances of heavy elements in the emission-line gas to be
overall 1.5 $\times$ solar. However, the carbon and 
nitrogen abundances are 3.2 and 6 times solar, respectively. One
possibility is that an early period of star-formation produced much of the
excess carbon, which was followed by conversion of carbon into nitrogen
in a more recent period. \citet{kallman2014} also suggested
non-standard abundances for the emission-line gas, but did not
discuss possible connections with the star-formation history.

3. We were able to fit most of strong emission lines with two components,
LOWION and HIGHION, characterized by log$U = -0.05$ and 1.22, and log$N_{\rm H} =
20.85$ and 21.2, respectively. The LOWION produces most of the emission
from He-like C, N, and O lines, while the HIGHION produces
the H-like N, O, and Ne and most of the He-like Ne lines. The predicted
electron temperature for LOWION is consistent with the measured widths of
He-like RRCs, however that of HIGHION is several times higher than that derived
from the H-like RRCs, which we attribute to uncertainities in the width 
measurements. Overall, the emission lines and RRCs are consistent with
photoionization, albeit with a small contribution from photoexcitation
for the resonance lines.

4. The covering factors determined for LOWION and HIGHION were 0.35 and 0.84, 
which, while physically possible, are high given the likelihood that a fraction
of the X-ray emission is undetected due to absorption by material in the disk
of the host galaxy or surrounding the NLR. However, the transparency of
these components to the ionizing radiation leads to the possibility that
there are multiple zones of highly ionized gas which subtend the same solid
angle. The additional emission from these zones would reduce the 
covering factors. 

5. Our estimated total mass and mass outflow rates for LOWION and HIGHION
are similar to the values determined by \citet{kallman2014} from the {\it Chandra}/HETG 
spectra. Interestingly, the X-ray emission line gas likely represents the
largest mass component of the NLR, which does not appear to be true of
NGC 4151 \citep[e.g.][]{armentrout2007}. 

6. The ionization state and temperature of HIGHION are consistent with those
of the scattering medium in which the polarized emission arises \citep[]{miller1991,
kraemer&crenshaw2000C}, although its column density is an order of magnitude too
small. However, if there are multiple zones, as noted in
item 4., the total column density of high-ionization gas could be sufficient
to produce the scattered light. The radial profile of the continuum radiation
\citep{crenshaw&kraemer2000A} is consistent with such a scenario. Therefore,
we suggest that the scattered emission arises in the X-ray emission-line
gas, in agreement with \citet{ogle2003}. 

Finally, the overall properties of NGC~1068, including a high mass accretion rate,
super-solar abundances, large amounts of highly ionized gas and molecular gas
\citep[e.g.][]{riffel2014}, and active star-formation
\citep{bruhweiler2001} may be connected to its stage of activity. That is, NGC~1068
is in an early part of its active phase, at which time the AGN is being
rapidly fueled, but before
the inner nucleus has been cleared.

\section{Acknowledgments}

This work was supported by NASA grant NNX10AD78G. 
This research has made use of the NASA/IPAC Extragalactic Database 
(NED), which is operated by the Jet Propulsion Laboratory, California 
Institute of Technology, under contract with the National Aeronautics 
and Space Administration. We thank the referee, Tim Kallman, for
his careful review of the paper and constructive comments. We thank A. Maeder, M. Catalan, and E.
Behar for useful suggestions. We also thank G. Ferland and associates
for their continuing maintenance of Cloudy.

\bibliographystyle{apj}      % basic style, author-year citations
\bibliography{ms}

\begin{thebibliography}{}
\expandafter\ifx\csname natexlab\endcsname\relax\def\natexlab#1{#1}\fi

\bibitem[{{Antonucci}(1993)}]{antonucci1993}
{Antonucci}, R.~R.~J. 1993, \araa, 31, 473

\bibitem[{{Armentrout} {et~al.}(2007){Armentrout}, {Kraemer}, \&
  {Turner}}]{armentrout2007}
{Armentrout}, B., {Kraemer}, S., \& {Turner}, T. 2007, \apj, 665, 237

\bibitem[{{Arnaud}(2010)}]{arnaud2010}
{Arnaud}, K.~A. 2010, in Bulletin of the American Astronomical Society,
  Vol.~42, AAS/High Energy Astrophysics Division \#11, 668

\bibitem[{{Asplund} {et~al.}(2005){Asplund}, {Grevesse}, \&
  {Sauval}}]{asplund2005}
{Asplund}, M., {Grevesse}, N., \& {Sauval}, A.~J. 2005, in Astronomical Society
  of the Pacific Conference Series, Vol. 336, Cosmic Abundances as Records of
  Stellar Evolution and Nucleosynthesis, ed. T.~G. {Barnes}, III \& F.~N.
  {Bash}, 25

\bibitem[{{Awaki} {et~al.}(2001){Awaki}, {Terashima}, {Hayashida}, \&
  {Sakano}}]{awaki2001}
{Awaki}, H., {Terashima}, Y., {Hayashida}, K., \& {Sakano}, M. 2001, PASP, 53,
  647

\bibitem[{{Bland-Hawthorn} {et~al.}(1997){Bland-Hawthorn}, {Gallimore},
  {Tacconi}, {Brinks}, {Baum}, {Antonucci}, \& {Cecil}}]{bland1997}
{Bland-Hawthorn}, J., {Gallimore}, J.~F., {Tacconi}, L.~J., {et~al.} 1997,
  \apss, 248, 9

\bibitem[{{Blustin} {et~al.}(2005){Blustin}, {Page}, {Fuerst},
  {Branduardi-Raymont}, \& {Ashton}}]{blustin2005}
{Blustin}, A.~J., {Page}, M.~J., {Fuerst}, S.~V., {Branduardi-Raymont}, G., \&
  {Ashton}, C.~E. 2005, A\&A, 431, 111

\bibitem[{{Bonatto} \& {Pastoriza}(1997)}]{bonatto1997}
{Bonatto}, C.~J., \& {Pastoriza}, M.~G. 1997, \apj, 486, 132

\bibitem[{{Bottorff} {et~al.}(2000){Bottorff}, {Korista}, \&
  {Shlosman}}]{bottorff2000}
{Bottorff}, M.~C., {Korista}, K.~T., \& {Shlosman}, I. 2000, \apj, 537, 134

\bibitem[{{Brinkman} {et~al.}(2002){Brinkman}, {Kaastra}, {van der Meer},
  {Kinkhabwala}, {Behar}, {Kahn}, {Paerels}, \& {Sako}}]{brinkman2002}
{Brinkman}, A.~C., {Kaastra}, J.~S., {van der Meer}, R. L.~J., {et~al.} 2002,
  A\&A, 396, 761

\bibitem[{{Bruhweiler} {et~al.}(2001){Bruhweiler}, {Miskey}, {Smith},
  {Landsman}, \& {Malumuth}}]{bruhweiler2001}
{Bruhweiler}, F.~C., {Miskey}, C.~L., {Smith}, A.~M., {Landsman}, W., \&
  {Malumuth}, E. 2001, \apj, 546, 866

\bibitem[{{Capetti} {et~al.}(1997){Capetti}, {Axon}, \&
  {Macchetto}}]{capetti1997}
{Capetti}, A., {Axon}, D.~J., \& {Macchetto}, F.~D. 1997, \apj, 487, 560

\bibitem[{{Carigi} {et~al.}(2005){Carigi}, {Peimbert}, {Esteban}, \&
  {García-Rojas}}]{carigi2005}
{Carigi}, L., {Peimbert}, M., {Esteban}, C., \& {García-Rojas}, J. 2005, \apj,
  623, 213

\bibitem[{{Crenshaw} \& {Kraemer}(2000{\natexlab{a}})}]{crenshaw&kraemer2000A}
{Crenshaw}, D.~M., \& {Kraemer}, S.~B. 2000{\natexlab{a}}, \apj, 532, 247

\bibitem[{{Crenshaw} \& {Kraemer}(2000{\natexlab{b}})}]{crenshaw&kraemer2000}
---. 2000{\natexlab{b}}, \apjl, 532, L101

\bibitem[{{Crenshaw} \& {Kraemer}(2012)}]{crenshaw2012}
---. 2012, \apj, 753, 75

\bibitem[{{Crenshaw} {et~al.}(2003){Crenshaw}, {Kraemer}, \&
  {George}}]{crenshaw2003}
{Crenshaw}, D.~M., {Kraemer}, S.~B., \& {George}, I. 2003, \araa, 41, 117

\bibitem[{{Crenshaw} {et~al.}(2010{\natexlab{a}}){Crenshaw}, {Kraemer},
  {Schmitt}, {Jaff\'{e}}, {Deo}, {Collins}, \& {Fischer}}]{crenshaw2010B}
{Crenshaw}, D.~M., {Kraemer}, S.~B., {Schmitt}, H.~R., {et~al.}
  2010{\natexlab{a}}, \aj, 139, 871

\bibitem[{{Crenshaw} {et~al.}(2010{\natexlab{b}}){Crenshaw}, {Schmitt},
  {Kraemer}, {Mushotzky}, \& {Dunn}}]{crenshaw2010}
{Crenshaw}, D.~M., {Schmitt}, H.~R., {Kraemer}, S.~B., {Mushotzky}, R.~F., \&
  {Dunn}, J.~P. 2010{\natexlab{b}}, \apj, 708, 419

\bibitem[{{Das} {et~al.}(2007){Das}, {Crenshaw}, \& {Kraemer}}]{vdas2007}
{Das}, V., {Crenshaw}, D.~M., \& {Kraemer}, S.~B. 2007, \apj, 656, 699

\bibitem[{{Das} {et~al.}(2006){Das}, {Crenshaw}, {Kraemer}, \&
  {Deo}}]{vdas2006}
{Das}, V., {Crenshaw}, D.~M., {Kraemer}, S.~B., \& {Deo}, R.~P. 2006, \aj, 132,
  620

\bibitem[{{Dickey} \& {Lockman}(1990)}]{dickey1990}
{Dickey}, J.~M., \& {Lockman}, F.~J. 1990, \araa, 28, 215

\bibitem[{{Edelson} {et~al.}(1996){Edelson}, {Alexander}, {Crenshaw}, {Kaspi},
  {Malkan}, {Peterson}, {Warwick}, {Clavel}, {Filippenko}, {Horne}, \&
  et~al.}]{edelson1996}
{Edelson}, R.~A., {Alexander}, T., {Crenshaw}, D.~M., {et~al.} 1996, \apj, 470,
  364

\bibitem[{{Esteban} {et~al.}(2005){Esteban}, {García-Rojas}, {Peimbert},
  {Peimbert}, {Ruiz}, {Rodríguez}, \& {Carigi}}]{esteban2005}
{Esteban}, C., {García-Rojas}, J., {Peimbert}, M., {et~al.} 2005, \apjl, 618,
  L95

\bibitem[{{Evans} {et~al.}(1991){Evans}, {Ford}, {Kinney}, {Antonucci},
  {Armus}, \& {Caganoff}}]{evans1991}
{Evans}, I.~N., {Ford}, H.~C., {Kinney}, A.~L., {et~al.} 1991, \apjl, 369, L27

\bibitem[{{Everett} \& {Murray}(2007)}]{everett2007}
{Everett}, J.~E., \& {Murray}, N. 2007, \apj, 656, 93

\bibitem[{{Ferland} {et~al.}(1998){Ferland}, {Korista}, {Verner}, {Ferguson},
  {Kingdon}, \& {Verner}}]{ferland1998}
{Ferland}, G.~J., {Korista}, K.~T., {Verner}, D.~A., {et~al.} 1998, \pasp, 110,
  761

\bibitem[{{Ferland} {et~al.}(2013){Ferland}, {Porter}, {van Hoof}, {Williams},
  {Abel}, {Lykins}, {Shaw}, {Henney}, \& {Stancil}}]{ferland2013}
{Ferland}, G.~J., {Porter}, R.~L., {van Hoof}, P.~A.~M., {et~al.} 2013, \rmxaa,
  49, 137

\bibitem[{{Gallimore} {et~al.}(1994){Gallimore}, {Baum}, {O'Dea}, {Brinks}, \&
  {Pedlar}}]{gallimore1994}
{Gallimore}, J.~F., {Baum}, S.~A., {O'Dea}, C.~P., {Brinks}, E., \& {Pedlar},
  A. 1994, \apjl, 422, L13

\bibitem[{{George} {et~al.}(1998){George}, {Turner}, {Mushotzky}, {Nandra}, \&
  {Netzer}}]{george1998}
{George}, I.~M., {Turner}, T.~J., {Mushotzky}, R., {Nandra}, K., \& {Netzer},
  H. 1998, \apjs, 114, 73

\bibitem[{{Greenhill} \& {Gwinn}(1997)}]{greenhill1997}
{Greenhill}, L.~J., \& {Gwinn}, C.~R. 1997, Ap\&SS, 248, 261

\bibitem[{{Grevesse} \& {Sauval}(1998)}]{grevesse1998}
{Grevesse}, N., \& {Sauval}, A.~J. 1998, \ssr, 85, 161

\bibitem[{{Halpern}(1984)}]{halpern1984}
{Halpern}, J. 1984, \apj, 281, 90

\bibitem[{{Henry} {et~al.}(2000){Henry}, {Edmunds}, \& {Köppen}}]{henry2000}
{Henry}, R. B.~C., {Edmunds}, M.~G., \& {Köppen}, J. 2000, \apj, 541, 660

\bibitem[{{Huchra} {et~al.}(1999){Huchra}, {Vogeley}, \& {Geller}}]{huchra1999}
{Huchra}, J.~P., {Vogeley}, M.~S., \& {Geller}, M.~J. 1999, VizieR Online Data
  Catalog, 212, 10287

\bibitem[{{Kalberla} {et~al.}(2005){Kalberla}, {Burton}, {Hartmann}, {Arnal},
  {Bajaja}, {Morras}, \& {P{\"o}ppel}}]{kalberla2005}
{Kalberla}, P.~M.~W., {Burton}, W.~B., {Hartmann}, D., {et~al.} 2005, \aap,
  440, 775

\bibitem[{{Kallman} {et~al.}(2014){Kallman}, {Evans}, {Marshall}, {Canizares},
  {Longinotti}, {Nowak}, \& {Schulz}}]{kallman2014}
{Kallman}, T., {Evans}, D.~A., {Marshall}, H., {et~al.} 2014, \apj, 780, 121

\bibitem[{{Kallman} {et~al.}(2004){Kallman}, {Palmeri}, {Bautista}, {Mendoza},
  \& {Krolik}}]{kallman2004}
{Kallman}, T.~R., {Palmeri}, P., {Bautista}, M.~A., {Mendoza}, C., \& {Krolik},
  J.~H. 2004, \apjs, 155, 675

\bibitem[{{Kinkhabwala} {et~al.}(2002){Kinkhabwala}, {Sako}, {Behar}, {Kahn},
  {Paerels}, {Brinkman}, {Kaastra}, {Gu}, \& {Liedahl}}]{kinkhabwala2002}
{Kinkhabwala}, A., {Sako}, M., {Behar}, E., {et~al.} 2002, \apj, 575, 732

\bibitem[{{Koski}(1978)}]{koski1978}
{Koski}, A.~T. 1978, \apj, 223, 56

\bibitem[{{Kraemer} \& {Crenshaw}(2000{\natexlab{a}})}]{kraemer&crenshaw2000C}
{Kraemer}, S.~B., \& {Crenshaw}, D.~M. 2000{\natexlab{a}}, \apj, 532, 256

\bibitem[{{Kraemer} \& {Crenshaw}(2000{\natexlab{b}})}]{kraemer&crenshaw2000}
---. 2000{\natexlab{b}}, \apj, 544, 763

\bibitem[{{Kraemer} {et~al.}(1998){Kraemer}, {Ruiz}, \& {Crenshaw}}]{krc1998}
{Kraemer}, S.~B., {Ruiz}, J.~R., \& {Crenshaw}, D.~M. 1998, \apj, 508, 232

\bibitem[{{Kraemer} {et~al.}(2011){Kraemer}, {Schmitt}, {Crenshaw},
  {Mel\'{e}ndez}, {Turner}, {Guainazzi}, \& {Mushotzky}}]{kraemer2011}
{Kraemer}, S.~B., {Schmitt}, H.~R., {Crenshaw}, D.~M., {et~al.} 2011, \apj,
  727, 130

\bibitem[{{Kraemer} {et~al.}(2009){Kraemer}, {Trippe}, {Crenshaw},
  {Mel\'{e}ndez}, {Schmitt}, \& {Fischer}}]{kraemer2009}
{Kraemer}, S.~B., {Trippe}, M.~L., {Crenshaw}, D.~M., {et~al.} 2009, \apj, 698,
  106

\bibitem[{{Kriss} {et~al.}(1992){Kriss}, {Davidsen}, {Blair}, {Ferguson}, \&
  {Long}}]{kriss1992}
{Kriss}, G.~A., {Davidsen}, A.~F., {Blair}, W.~P., {Ferguson}, H.~C., \&
  {Long}, K.~S. 1992, \apjl, 394, L37

\bibitem[{{Krolik} \& {Kriss}(2001)}]{krolik2001}
{Krolik}, J.~H., \& {Kriss}, G.~A. 2001, \apj, 561, 684

\bibitem[{{Lutz} {et~al.}(2000){Lutz}, {Sturm}, {Genzel}, {Moorwood},
  {Alexander}, {Netzer}, \& {Sternberg}}]{lutz2000}
{Lutz}, D., {Sturm}, E., {Genzel}, R., {et~al.} 2000, \apj, 536, 697

\bibitem[{{Maeder} \& {Meynet}(1989)}]{maeder1989}
{Maeder}, A., \& {Meynet}, G. 1989, A\&A, 210, 155

\bibitem[{{Mel\'{e}ndez} {et~al.}(2008){Mel\'{e}ndez}, {Kraemer}, {Deo},
  {Crenshaw}, {Schmitt}, {Mushotzky}, {Tueller}, {Markwardt}, \&
  {Winter}}]{melendez2008}
{Mel\'{e}ndez}, M., {Kraemer}, S. B. amd~{Armentrout}, B.~K., {Deo}, R.~P.,
  {et~al.} 2008, \apj, 682, 94

\bibitem[{{Miller} {et~al.}(1991){Miller}, {Goodrich}, \&
  {Mathews}}]{miller1991}
{Miller}, J.~S., {Goodrich}, R., \& {Mathews}, W.~G. 1991, \apj, 378, 47

\bibitem[{{Netzer}(1993)}]{netzer1993}
{Netzer}, H. 1993, \apj, 411, 594

\bibitem[{{Netzer}(1996)}]{netzer1996}
---. 1996, \apj, 473, 781

\bibitem[{{Ogle} {et~al.}(2003){Ogle}, {Brookings}, {Canizares}, {Lee}, \&
  {Marshall}}]{ogle2003}
{Ogle}, P.~M., {Brookings}, T., {Canizares}, C.~R., {Lee}, J.~C., \&
  {Marshall}, H.~L. 2003, \aap, 402, 849

\bibitem[{{Ogle} {et~al.}(2000){Ogle}, {Marshall}, {Lee}, \&
  {Canizares}}]{ogle2000}
{Ogle}, P.~M., {Marshall}, H.~L., {Lee}, J.~C., \& {Canizares}, C.~R. 2000,
  \apjl, 545, L81

\bibitem[{{Peimbert}(1967)}]{peimbert1967}
{Peimbert}, M. 1967, \apj, 150, 825

\bibitem[{{Peterson} {et~al.}(2004){Peterson}, {Ferrarese}, {Gilbert}, {Kaspi},
  {Malkan}, {Maoz}, {Merritt}, {Netzer}, \& {Onken}}]{peterson2004}
{Peterson}, B.~M., {Ferrarese}, L., {Gilbert}, K.~M., {et~al.} 2004, \apj, 613,
  682

\bibitem[{{Pier} {et~al.}(1994){Pier}, {Antonucci}, {Hurt}, {Kriss}, \&
  {Krolik}}]{pier1994}
{Pier}, E.~A., {Antonucci}, R., {Hurt}, T., {Kriss}, G., \& {Krolik}, J. 1994,
  \apj, 428, 124

\bibitem[{{Pogge}(1988)}]{pogge1988}
{Pogge}, R.~W. 1988, \apj, 328, 519

\bibitem[{{Porquet} \& {Dubau}(2002)}]{porquet&dubau2000}
{Porquet}, D., \& {Dubau}, J. 2002, in SF2A-2002: Semaine de l'Astrophysique
  Francaise, ed. F.~{Combes} \& D.~{Barret}, 371

\bibitem[{{Porter} {et~al.}(2006){Porter}, {Ferland}, {Kraemer}, {Armentrout},
  {Arnaud}, \& {Turner}}]{porter2006}
{Porter}, R.~L., {Ferland}, G.~J., {Kraemer}, S.~B., {et~al.} 2006, \pasp, 118,
  920

\bibitem[{{Riffel} {et~al.}(2014){Riffel}, {Vale}, {Storchi-Bergmann}, \&
  {McGregor}}]{riffel2014}
{Riffel}, R.~A., {Vale}, T.~B., {Storchi-Bergmann}, T., \& {McGregor}, P.~J.
  2014, \mnras, 442, 656

\bibitem[{{Seaton}(1978)}]{seaton1978}
{Seaton}, M.~J. 1978, \mnras, 185, 5P

\bibitem[{{Seaton}(1979)}]{seaton1979}
---. 1979, \mnras, 187, 73P

\bibitem[{{Sofue}(1991)}]{sofue1991}
{Sofue}, Y. 1991, \pasj, 43, 671

\bibitem[{{Tully} {et~al.}(2009){Tully}, {Rizzi}, {Shaya}, {Courtois},
  {Makarov}, \& {Jacobs}}]{tully2009}
{Tully}, R., {Rizzi}, L., {Shaya}, E., {et~al.} 2009, \aj, 138, 323

\bibitem[{{Turner} {et~al.}(2003){Turner}, {Kraemer}, {Mushotzky}, {George}, \&
  {Gabel}}]{turner2003}
{Turner}, T.~J., {Kraemer}, S.~B., {Mushotzky}, R.~F., {George}, I.~M., \&
  {Gabel}, J.~R. 2003, \apj, 594, 128

\bibitem[{{Whittle}(1992)}]{whittle1992}
{Whittle}, M. 1992, \apjs, 79, 49

\bibitem[{{Wilms} {et~al.}(2000){Wilms}, {Allen}, \& {McCray}}]{wilms2000}
{Wilms}, J., {Allen}, A., \& {McCray}, R. 2000, \apj, 542, 914

\bibitem[{{Young} {et~al.}(2001){Young}, {Wilson}, \& {Shopbell}}]{young2001}
{Young}, A.~J., {Wilson}, A.~S., \& {Shopbell}, P.~L. 2001, \apj, 556, 6

\end{thebibliography}

%\bibliography{ngc1068_paper1}   % name your BibTeX data base

%%%%%%%%%%%%%%%%%%%%%%%%%%%%%%%%%%%%%%%%%%%%%%%%%%%
%%%%%%%%%%%%%%%%%%%%%%%%%%%%%%%%%%%%%%%%%%%%%%%%%%%
%%%%%%%%%%%%%%%%%%%%%%%%%%%%%%%%%%%%%%%%%%%%%%%%%%%
%%%%%%%%%%%%%%%%%%%%%%%%%%%%%%%%%%%%%%%%%%%%%%%%%%%

\begin{center}
\begin{table}[ht]
\footnotesize
\vspace{-0.3in}

 \setlength{\tabcolsep}{15pt}

 \caption{Comparison of the Observed and Predicted Emission-Line Fluxes}
  
 \vspace{0.2in}
 
 \begin{threeparttable}
 
 \begin{tabular}{ p{2.9cm} p{2.1cm} p{2.5cm} p{2.1cm} p{1.5cm} p{1.5cm}} \hline \hline

Line ID & 			Observed\tnote{a} & Absorption-corrected    & Total Model\tnote{b} & Low U\tnote{c} & High U\tnote{d}  \\ \hline \hline

C {\sc v} He${\beta}$            &        1.16${\pm}$0.31                        &   3.24 ${\pm}$ 0.86       &  1.09		& 1.09      	&	  \\
C {\sc v} He${\delta}$           &        2.45 ${\pm}$ 1.30                      &  4.62${\pm2.08}$ 	&   0.59		& 0.59       	&  \\
C {\sc vi} Ly${\alpha}$          &       9.48${\pm}$ 0.41           	 	&  23.35${\pm}$ 1.01    	 &  16.57		 & 9.55	  & 7.02 \\
C {\sc vi} Ly${\beta}$            &        1.76 ${\pm}$ 0.21            	&  3.18${\pm}$ 0.38      	 &    2.58		& 0.87      	& 1.71 \\
C {\sc vi} Ly${\gamma}$       &        0.60${\pm}$ 0.14        		&  0.95${\pm}$ 0.22      	&     1.41 		& 0.53 	&  0.88 \\
C {\sc vi} Ly${\delta}$            &       1.07${\pm}$ 0.16        		&  1.94${\pm}$ 0.29      	&     0.87 		& 0.41  	& 0.46 \\

N {\sc vi} He${\gamma}$        &  0.47 ${\pm}$  0.14                      &   0.67 ${\pm}$ 0.20           &  0.42		& 0.42        &   \\
N {\sc vi} {\it r}                           &  3.32${\pm}$  0.48                        &   6.08 ${\pm}$ 0.87       &  4.13 		& 3.62       	& 0.51 \\
N {\sc vi} {\it i}                           &  0.93${\pm}$  0.37                          &   1.72 ${\pm}$ 0.68           &   1.80		& 1.80  &   \\
N {\sc vi} {\it f}                           &  7.37 ${\pm}$  0.12                         &   13.20 ${\pm}$ 0.60          &   9.77	& 9.77       &   \\
N {\sc vii} Ly${\alpha}$           &  6.01 ${\pm}$  0.26                         &   8.92 ${\pm}$ 0.39           &   6.45 	&  1.99      & 4.46 \\
N {\sc vii} Ly${\beta}$              & 0.95${\pm}$  0.09                           &   1.45 ${\pm}$ 0.16            &    1.62 	& 0.43   	 & 1.19 \\
N {\sc vii} Ly${\gamma}$        &  0.40 ${\pm}$  0.16                         & 0.53 ${\pm}$ 0.16              &    0.88	& 0.27      	&  0.61 \\
N {\sc vii} Ly${\delta}$             &  0.35${\pm}$  0.10                         &  0.40 ${\pm}$ 0.10            &    0.51 	& 0.19    &  0.32 \\

O {\sc vii} {\it r}                      & 4.96 ${\pm}$  0.28                                &   7.85 ${\pm}$ 0.68            & 5.30 	& 3.97    	& 1.33 \\
O {\sc vii} {\it i}                        & 1.00 ${\pm}$  0.44                               &  1.94 ${\pm}$ 0.70      	 & 3.13	& 3.13   &  \\
O {\sc vii} {\it f}                         &   9.25 ${\pm}$ 0.18                             &   14.99 ${\pm}$ 0.29          & 12.57 	& 12.07  & 0.50  \\
O {\sc vii} He${\beta}$            &   0.73$\pm$ 0.15                           &    0.99 ${\pm}$ 0.22          & 0.88 	& 0.54         &  0.34  \\
O {\sc vii} He${\gamma}$       &  0.67 ${\pm}$     0.09                         & 0.88 ${\pm}$ 0.14               & 0.36	& 0.36     &  \\
O {\sc vii} He${\delta}$            &  0.38${\pm}$  0.08                             & 0.49${\pm}$ 0.12              & 0.30 	& 0.30            &  \\
O {\sc viii} Ly${\alpha}$           &   5.37${\pm}$  0.30                           & 7.44${\pm}$ 0.64             & 8.94	& 1.25               & 7.69 \\    
O {\sc viii} Ly${\beta}$              &  0.96  $\pm$ 0.44               	       & 1.22 ${\pm}$ 0.28           & 1.59  	& 0.32               & 1.27 \\

Ne {\sc ix} {\it r}                            &  1.74 $\pm$0.13                                  &  2.08 ${\pm}$ 0.15           &  1.46  	& 0.45  & 1.01  \\
Ne {\sc ix} {\it f}                            &  1.11  $\pm$ 0.15                      &   1.32 ${\pm}$ 0.18           	 & 0.79 	& 0.38  &   0.41  \\
Ne {\sc ix} He${\beta}$              &  0.21   $\pm$ 0.09                     &   0.24 ${\pm}$ 0.09            	& 0.45 	&  0.16         &   0.29  \\
Ne {\sc x} Ly${\alpha}$             &  1.34 ${\pm}$  0.11                    & 1.52 ${\pm}$ 0.12              	& 2.07 	&  		&  2.07 \\ 
Mg {\sc xi} {\it rif}                         &  0.90 $\pm$ 0.17                       & 0.96 ${\pm}$ 0.19              	& 1.46 	&       &   1.46 \\
%Mg {\sc xi} ${\beta}$                  &  . . . . . . .                                 		& ${<}$ 0.36             		& . . .                                 & 0.30  \\
Mg {\sc xii} Ly${\alpha}$          & 0.27   $\pm$0.11                        		& 0.29 ${\pm}$ 0.12             &  0.95	&   &  0.95 \\   

\label{observations}
\end{tabular}

\begin{tablenotes}
\item[a]{Fluxes obtained by fitting Gaussian to each line as identified in the combined spectra of RGS 1 \& 2.;
all fluxes $\times$ 10${^{-4}}$ photons cm${^{-2}}$ s${^{-1}}$.}
\item[a]{Absorption-corrected line fluxes predicted by {\sc cloudy} in  the total model. }
\item[b]{Fluxes predicted by {\sc cloudy} for LOWION, after scaling to that of O~{\sc vii} {\it f}.}
\item[c]{Fluxes predicted by {\sc cloudy} for HIGHION, after scaling to that of O~{\sc viii} Ly$\alpha$.}

\end{tablenotes}

\end{threeparttable}

\end{table}

\begin{table}[ht]
\footnotesize
\setlength{\tabcolsep}{12pt}

 \caption{Observed Energies and Velocity Shifts Relative to Systemic} 
 \vspace{0.2in}

\begin{threeparttable}

 \begin{tabular}{  p{1.8cm} p{1.8cm} p{1.8cm} p{2.5cm} p{3.5cm}  } \hline \hline

Line ID & ${\lambda}_{\rm th}$\tnote{a}  & $E_{\rm th}$\tnote{b}   & $\Delta E$\tnote{c} & Velocity shift\tnote{d}    \\ 

&  ({\AA}) & (eV) & (eV) &(km s${^{-1}}$) \\ \hline \hline

{\rm C} {\sc v} He${\beta}$     &  34.9728         & 354.516          & 0.444 ${\pm}$ 0.085              & -490 ${\pm}$ 80  \\
C {\sc v} He${\delta}$     &  32.7542                 &378.529          & 0.969 ${\pm}$ 0.205               & -680 ${\pm}$ 120\\
C {\sc vi} Ly${\alpha}$     &  33.7342                 & 367.533          & 0.333 ${\pm}$ 0.019               & -270 ${\pm}$ 50 \\
C {\sc vi} Ly${\beta}$          &28.4663                & 435.547          & 0.582 ${\pm}$ 0.065               & -400 ${\pm}$ 70 \\
C {\sc vi} Ly${\gamma}$       &  26.9900                & 459.369          & 0.618 ${\pm}$ 0.040              & -400 ${\pm}$ 80 \\
C {\sc vi} Ly${\delta}$         & 26.3572                & 470.399          & 0.459 ${\pm}$ 0.135               & -290 ${\pm}$ 10\\

N {\sc vi} He${\gamma}$    & 23.7710                & 521.578          &0.685 ${\pm}$ 0.161              & -400 ${\pm}$ 110 \\
N {\sc vi} r                        & 28.7870                 & 430.695          & 0.376 ${\pm}$ 0.042               & -260 ${\pm}$ 60 \\
N {\sc vi} i                       & 29.0815                 &426.333           &  0.342 ${\pm}$ 0.069               & -240 ${\pm}$ 70 \\
N {\sc vi} f                        &29.5343                 &419.797           &  0.495 ${\pm}$ 0.018               & -350 ${\pm}$ 60\\
N {\sc vii} Ly${\alpha}$     &24.7792                 & 500.356          & 0.318 ${\pm}$ 0.045                & -190 ${\pm}$ 70\\
N {\sc vii} Ly${\beta}$       &20.9095                 & 592.957          &0.498
 ${\pm}$ 0.141                & -251 ${\pm}$ 102\\
N {\sc vii} Ly${\gamma}$    & 19.8261                 & 625.358          & 0.513 ${\pm}$ 0.161                & -250 ${\pm}$ 110\\
N {\sc vii} Ly${\delta}$      &19.3614                 & 640.368         & 0.841 ${\pm}$ 0.235                & -390 ${\pm}$ 130\\

O {\sc vii} r                   & 21.6020                 &573.947           &   0.483 ${\pm}$ 0.046               & -250 ${\pm}$ 80 \\
O {\sc vii} i                     & 21.8044                &568.620            &  0.539 ${\pm}$ 0.111              & -280 ${\pm}$ 90\\
O {\sc vii} f                     & 22.1012                 &560.983           & 0.708 ${\pm}$ 0.029                 &  -380 ${\pm}$ 70\\
O {\sc vii} He${\beta}$     & 18.6270                &665.615            & 1.516 ${\pm}$ 0.174                 & -680${\pm}$ 120 \\
O {\sc vii} He${\gamma}$     & 17.7682                &697.787            & 1.414 ${\pm}$ 0.221                 & -610 ${\pm}$ 130\\
O {\sc vii} He${\delta}$     & 17.3958                &712.722          &1.145 ${\pm}$ 0.310                 & -480 ${\pm}$ 150 \\
O {\sc viii} Ly${\alpha}$   &18.9725                 & 653.493           & 0.337 ${\pm}$ 0.036                 & -150 ${\pm}$ 80\\ 
O {\sc viii} Ly${\beta}$   &16.0067                 & 774.577         &-0.111 ${\pm}$ 0.194                 & +40 ${\pm}$ 120\\ 
%Ne {\sc ix} r                      & 13.4470              &922.016          & 0.571  ${\pm}$ 0.287                          & +260 ${\pm}$ 133 \\ 
%Ne {\sc ix} f                        & 13.6970               & 905.192           & 0.593  ${\pm}$ 0.223                & +104 ${\pm}$ 148 \\ 
%Ne {\sc x} Ly${\alpha}$   & 12.1339                 & 1021.80           &-0.300 ${\pm}$ 0.206               & +88 ${\pm}$ 142\\ 
%Fe {\sc xvii}    & 15.0140                 & 825.700           &-0.318 ${\pm}$ 0.147               & +116 ${\pm}$ 116\\ 
%Fe {\sc xvii}    & 17.0510                 & 727.138           &-0.369 ${\pm}$ 0.152               & +152 ${\pm}$ 105\\  
%Fe {\sc xxiv}    & 10.6190                 & 1167.57           &-0.930 ${\pm}$ 0.929               & +239 ${\pm}$ 284 \\  \hline

\label{velocity shift}
\end{tabular}

\begin{tablenotes}

\item[a]{Theoretical wavelengths from NIST/Kentucky atomic database.}
\item[b]{Theoretical line energies.}
\item[c]{$\Delta E$=$E_{\rm obs}$-$E_{\rm th}$,  with uncertainties of 1${\sigma}$ (68\% confidence level).}
\item[d]{The quoted errors are statistical (1${\sigma}$) and systematic (derived from the centroid position uncertainity), combined in quadrature.}

\end{tablenotes}

\end{threeparttable}

\end{table}

\begin{deluxetable}{lccc}
\tablecolumns{4}
\footnotesize
\tablecaption{Photoionization Model Parameters$^{a}$}
\tablewidth{0pt}
\tablehead{
\colhead{Component} & \colhead{logU} & \colhead{logN$_{\rm H}^{b}$}
& \colhead{velocity$^{c}$}  } 
\startdata
LOWION & $-$0.05 $^{+0.02}_{-0.02}$ & 20.85 $^{+0.03}_{-0.05}$ & 215 $^{+12}_{-22}$ \\
HIGHION & 1.23 $^{+0.01}_{-0.01}$ & 21.2 $^{+0.00}_{-0.001}$$^{d}$ & 166 $^{+23}_{-39}$    \\
\enddata

\tablenotetext{a}{Fit statistics: $\chi^{2}$ $= 9805.83$ using 4926 PHA bins.
Reduced $\chi^{2} = 1.99508$ for 4915 degrees of freedom.}
\tablenotetext{b}{N$_{\rm H}$ in units of cm$^{-2}$}
\tablenotetext{c}{Velocity offset (km s$^{-1}$) from systemic.}
\tablenotetext{d}{Upper limit fixed based on $\Delta R/R$ constraint
(see Section~\ref{Final Fit}).}

\label{model parameters}
\end{deluxetable}

\begin{deluxetable}{lccccc}
\tablecolumns{6}
\footnotesize
\tablecaption{Measured and Model-Predicted RRC Parameters}
\tablewidth{0pt}
\tablehead{
\colhead{Ion} & \colhead{kT (eV)} & \colhead{Flux$^{a}$} & \colhead{Model Total} 
&\colhead {LOWION} &\colhead{HIGHION}}
\startdata
C~{\sc v} & 2.39$\pm$0.50 & 11.25$\pm$0.90 & 6.51 & 6.51 & \\
C~{\sc vi} & 8.13$\pm$2.10 & 9.25$\pm$0.91 & 9.62 & 2.48 & 4.03 \\
N~{\sc vi} & 1.01$\pm$0.33 & 1.40$\pm$0.13 & 2.56 & 2.56 &  \\
N~{\sc vii} & 6.56$\pm$0.30 & 4.38$\pm$0.16 & 3.03 & 0.77 & 2.26 \\
O~{\sc vii} & 4.50$\pm$0.25 & 2.97$\pm$0.30 & 4.46 & 4.15 & 0.31 \\
\enddata

\tablenotetext{a}{All fluxes $\times$ 10${^{-4}}$ photons cm${^{-2}}$ s${^{-1}}$}

\label{rrcs}
\end{deluxetable}

\end{center}

\end{document}